\documentclass[showpacs,prb,twocolumn,floatfix]{revtex4}

\usepackage{graphicx}
\usepackage{amsmath}
\usepackage{amssymb}
\usepackage[colorlinks=true,citecolor=blue,linkcolor=blue]{hyperref}
\usepackage{amsfonts}
\usepackage{color,xcolor}
\usepackage{epstopdf}
\usepackage{braket}
\usepackage{bm}

\graphicspath{{figures/}}
\renewcommand{\vec}[1]{\ensuremath{\boldsymbol{#1}}}

\begin{document}

\title{Piezoelectricity in asymmetrically strained bilayer graphene}
\date{\today}
\author{M. Van der Donck}
\email{matthias.vanderdonck@uantwerpen.be}
\affiliation{Department of Physics, University of Antwerp, Groenenborgerlaan 171, B-2020 Antwerp, Belgium}
\author{C. De Beule}
\affiliation{Department of Physics, University of Antwerp, Groenenborgerlaan 171, B-2020 Antwerp, Belgium}
\author{B. Partoens}
\affiliation{Department of Physics, University of Antwerp, Groenenborgerlaan 171, B-2020 Antwerp, Belgium}
\author{F. M. Peeters}
\email{francois.peeters@uantwerpen.be}
\affiliation{Department of Physics, University of Antwerp, Groenenborgerlaan 171, B-2020 Antwerp, Belgium}
\author{B. Van Duppen}
\email{ben.vanduppen@uantwerpen.be}
\affiliation{Department of Physics, University of Antwerp, Groenenborgerlaan 171, B-2020 Antwerp, Belgium}

\begin{abstract}
We study the electronic properties of commensurate faulted bilayer graphene by diagonalizing the one-particle Hamiltonian of the bilayer system in a complete basis of Bloch states of the individual graphene layers. Our novel approach is very general and can be easily extended to any commensurate graphene-based heterostructure. Here, we consider three cases: i) twisted bilayer graphene, ii) bilayer graphene where triaxial stress is applied to one layer, and iii) bilayer graphene where uniaxial stress is applied to one layer. We show that the resulting superstructures can be divided into distinct classes, depending on the twist angle or the magnitude of the induced strain. The different classes are distinguished from each other by the interlayer coupling mechanism, resulting in fundamentally different low-energy physics. For the cases of triaxial and uniaxial stress, the individual graphene layers tend to decouple and we find significant charge transfer between the layers. In addition, this piezoelectric effect can be tuned by applying a perpendicular electric field. Finally, we show how our approach can be generalized to multilayer systems.
\end{abstract}

\maketitle

\section{Introduction}

Ever since the discovery of graphene \cite{ontdekking}, a lot of research has been devoted to its electronic properties \cite{elecpropgraph}. Subsequently, attention turned to bilayer graphene, a set of two graphene layers that are coupled via weak van der Waals forces \cite{bilaag}. Its electronic properties depend strongly on the stacking configuration. The two most common high-symmetry configurations, AA and AB (Bernal) stacking, have a drastiscally different energy spectrum \cite{ABAA1,ABAA2}. In realistic samples, however, this symmetry is easily broken as a consequence of mechanical forces acting on the sample, leading to faulted bilayer graphene. Furthermore, the increased control of the number of graphene layers in few-layer systems \cite{lagen1,lagen2} has created opportunities to engineer new types of deformations of the perfect bilayer system.

The deformation of any graphene system generally results in an additional periodicity with a length scale much larger than the nearest-neighbor interatomic distance. The best known example of such superstructures can be found in twisted bilayer graphene, where the two layers are rotated with respect to each other, leading to the appearance of moir\'e patterns \cite{moire}. Another example is given by bilayer graphene where mechanical stress is applied to only one of the layers. The electronic properties of these structures were, for example, studied with standard tight-binding \cite{rotflat,rotrus} or first-principle calculations \cite{rotdft,rotdft2}. When the layers are only slightly deformed, however, this can be a formidable task because the unit cell becomes increasingly larger for smaller deformations and one must resort to continuum models \cite{rotoer,rottopol,rotcont,rotkosh,blochexp,mele,mele2}. Also, these methods do not give much insight into the interlayer coupling mechanism between the two graphene layers. This is highly relevant for faulted bilayer graphene systems, since charge carriers reside in both layers, in contrast to heterostructures made from graphene and an insulating substrate such as hexagonal boron nitride \cite{hbn1,hbn2,hbn3,hbn4}. It is therefore advantageous to consider a different approach.

For twisted bilayer graphene, some effective models have been constructed \cite{rotoer,rottopol,rotcont,rotkosh} based on the assumption that, since one is often only interested in the low-energy physics, the electronic properties of the separate layers can be approximated by Dirac cones. These models are usually further limited to small twist angles for which intervalley coupling can be neglected \cite{rotoer,rotcont}. On the other hand, the case of uniaxial stress applied to one of the layers of bilayer graphene has not been studied in depth \cite{uni1,uni2}. Moreover, the case of triaxial stress applied to one of the layers has, to the best of our knowledge, not yet been investigated. In both systems where stress is applied to one layer, we find that charge can be transferred between the layers. Therefore, applying triaxial or uniaxial stress to one of the layers of bilayer graphene results in a piezoelectric effect. Recently, there has been a growing interest in bilayer systems that exhibit piezoelectricity, for example in asymmetrically doped twisted bilayer graphene \cite{tbgdope1,tbgdope2}, in graphene doped with surface atoms \cite{graphdope}, and in two-dimensional Mo$\text{S}_2$ \cite{mos2,mos22} and graphene nitride \cite{graphnitr}.

In this paper, we present a model for commensurate faulted bilayer graphene by diagonalizing the one-particle Hamiltonian in a complete basis of Bloch states of the individual graphene layers. This model naturally includes intervalley scattering between the Dirac cones of the individual layers and works for any commensurate twist angle or strain. Moreover, our model can be applied to any commensurate bilayer system, it provides a clear recipe for constructing a low-energy model, and it can be easily extended to multilayer systems.

The paper is organized as follows. In Sec.\ \ref{sec:General theory} we construct a general theory for commensurate faulted bilayer graphene. Subsequently, we consider twisted bilayer graphene in Sec.\ \ref{sec:Twisted bilayer graphene} and we show that our approach reproduces known results from the literature. Next, in Sec.\ \ref{sec:Triaxial strain} and Sec.\ \ref{sec:Uniaxial strain}, we consider bilayer graphene where one of the layers is subjected to triaxial or uniaxial stress, respectively. Finally, in Sec.\ \ref{sec:Multilayer} we show how our approach can be generalized to multilayer systems and in Sec.\ \ref{sec:Conclusions} we conclude by summarizing the main results of this paper.

\section{Theory} \label{sec:General theory}

Consider bilayer graphene subject to a mechanical manipulation that e.g.\ rotates or strains the top layer with respect to the bottom layer. Such a deformation generally leads to a commensurate superstructure. The total one-particle Hamiltonian for the $p_z$ electrons of a faulted bilayer system can be written as
\begin{equation}
\label{schrod}
\hat H = \hat H_0^b + \hat H_0^t + \hat U,
\end{equation}
where $\hat H_0^b$ and $\hat H_0^t$ are the Hamiltonian of the bottom and top layer, respectively, and $\hat U$ is the interlayer coupling. For commensurate structures, the one-particle wave function $\left| \Psi_{\bm k} \right>$ of the faulted bilayer is labeled by a Bloch momentum $\bm k$ that lies in the superlattice Brillouin zone (SBZ). The Schr\"odinger equation becomes
\begin{equation}
\left( \hat H_0^b + \hat H_0^t + \hat U \right) \left| \Psi_{\bm k} \right> = E_{\bm k} \left| \Psi_{\bm k} \right>,
\end{equation}
with $E_{\bm k}$ the energy eigenvalue. To proceed, we insert a complete basis of Bloch states $\left| \Phi_{\bm \kappa}^{i,\chi} \right>$ of the individual layers and project it onto $\bra{\Phi_{\bm k + \bm G}^{i,\chi}}$:
\begin{multline}
\label{schrodbloch}
\sum_{i',\chi'} \sum_{\bm \kappa' \in BZ^{(i')}} \braket{\Phi_{\bm k + \bm G}^{i,\chi} | \hat H_0^b + \hat H_0^t + \hat U | \Phi_{\bm \kappa'}^{i',\chi'}} \braket{\Phi_{\bm \kappa'}^{i',\chi'} | \Psi_{\bm k}} \\
= E_{\bm k} \braket{\Phi_{\bm k + \bm G}^{i,\chi} | \Psi_{\bm k}},
\end{multline}
where $i, i'= b, t$ is a layer index and $\chi, \chi' = A, B$ is a sublattice index. To satisfy Bloch's theorem, $\left| \Psi_{\bm k} \right>$ can only contain basis states at momenta $\bm k + \bm G$, where $\bm G$ is a reciprocal superlattice vector that lies inside the Brillouin zone (BZ) of the respective layer. The basis states are therefore labeled by four indices and are explicitly given by
\begin{equation}
\label{blochsub}
\ket{\Phi^{i,\chi}_{\bm k + \bm G}} = \frac{1}{\sqrt{N_i}} \sum_{\bm R_{\chi}^i} e^{i \left( \bm k + \bm G \right) \cdot \bm R_{\chi}^i} \ket{\varphi_{\bm R_{\chi}^i}},
\end{equation}
with $N_i$ the number of unit cells of the $i$th layer, $\bm R_\chi^i$ a lattice vector of sublattice $\chi$ in layer $i$, and $\ket{\varphi_{\bm R_\chi^i}}$ the $p_z$ state of the atom at $\bm R_\chi^i$ [\onlinecite{tightbinding}].

The matrix elements of the intralayer Hamiltonian are nonzero only for basis states with the same layer index and the same momentum. In the nearest-neighbor approximation, the intralayer matrix elements become \cite{elecpropgraph}
\begin{equation}
\label{H0}
\left[ H_0^i(\bm \kappa) \right]_{\chi,\chi'} \equiv \braket{\Phi_{\bm \kappa}^{i,\chi} | \hat H_0^i | \Phi_{\bm \kappa}^{i,\chi'}} = 
\begin{pmatrix} 0 & f_i(\bm \kappa) \\ f_i^*(\bm \kappa) & 0 \end{pmatrix},
\end{equation}
with $f_i(\bm \kappa) = \sum_{j=1}^3 \gamma_{0j}^ie^{i\bm \kappa \cdot \bm \delta_j^i}$, and where $\bm \delta_j^i$ are the three nearest-neighbor bond vectors of layer $i$ with $\gamma_{0j}^i$ the corresponding intralayer hopping parameters. These parameters are not isotropic in the case of uniaxial strain. For pristine graphene, we have $\gamma_{0j}^i = \gamma_0 = 3.12$ eV [\onlinecite{parameters}]. 

In order to calculate the interlayer coupling matrix elements, we first note that
\begin{equation}
\label{UGbew}
\begin{aligned}
\braket{\Phi^{i,\chi}_{\bm \kappa} | \hat U | \Phi^{i',\chi'}_{\bm \kappa'}} & = \braket{\Phi^{i,\chi}_{\bm \kappa} | \hat T_{\bm L}^\dag \hat U \hat T_{\bm L} | \Phi^{i',\chi'}_{\bm \kappa'}} \\
&= e^{i \bm L \cdot (\bm \kappa - \bm \kappa')} \braket{\Phi^{i,\chi}_{\bm \kappa} | \hat U | \Phi^{i',\chi'}_{\bm \kappa'}},
\end{aligned}
\end{equation}
where $\hat T_{\bm L}$ is the translation operator, $\bm L$ is a superlattice vector, and we used the fact that the interlayer coupling $\hat U$ has the same periodicity as the superlattice. It follows that  the interlayer coupling matrix elements vanish unless $\bm \kappa - \bm \kappa' = \bm G$, since $\bm G \cdot \bm L = 2\pi n$ with $n$ an integer. Consequently, we only need to calculate the matrix element
\begin{align}
\braket{\Phi^{b,\chi}_{\bm k + \bm G} | \hat U | \Phi^{t,\chi'}_{\bm k + \bm G'}} & = \frac{1}{\sqrt{N_b N_t}} \sum_{\bm R_\chi^b, \bm R_{\chi'}^t} e^{i(\bm G' - \bm G) \cdot \bm R_{\chi'}^t} \nonumber \\
\times & e^{i(\bm k + \bm G) \cdot (\bm R_{\chi'}^t - \bm R_\chi^b)} \braket{\varphi_{\bm R_\chi^b} | \hat U | \varphi_{\bm R_{\chi'}^t}} \label{matrixelemalg},
\end{align}
where $\bm G'$ is another reciprocal superlattice vector. We further assume that $\braket{\varphi_{\bm R_\chi^b} | \hat U | \varphi_{\bm R_{\chi'}^t}} \equiv U(|\bm R_\chi^b - \bm R_{\chi'}^t|)$ only depends on the distance between the atoms. Usually, the following ansatz is used to model this function \cite{rotoer,rotcont}:
\begin{equation}
\label{g1var}
U(r) = \gamma_1e^{-\alpha \left(\sqrt{1+(r/c)^2}-1\right)},
\end{equation}
where $r$ is the in-plane distance, $\gamma_1 = 0.377$~eV is the interlayer hopping parameter (for AB stacking) between atoms that lie on top of each other, and $c=3.35$ \AA\ is the interlayer distance \cite{parameters}.  The value of $\gamma_1$ in reality depends on the environment of the eclipsing atoms, but this only leads to small quantitative differences. We have chosen $\alpha$ such that $U(a_0)$, with $a_0 = 1.42$ \AA\ the interatomic distance \cite{elecpropgraf}, reduces to the interlayer skew hopping parameter $\gamma_4$ of pristine bilayer graphene. Taking $\gamma_3$ instead gives small quantitative differences. For $\gamma_4 = 0.12$ eV [\onlinecite{parameters}], we obtain $\alpha = 13.29$.

The sum in Eq. \eqref{matrixelemalg} can be simplified by noticing that the sum over $\bm R_\chi^b$ is unchanged for $\bm R_{\chi'}^t \rightarrow \bm R_{\chi'}^t + \bm L$ with $\bm L$ a superlattice vector. We obtain
\begin{equation}
\label{matrixelemalgfin}
\begin{aligned}
U^{\chi,\chi'}_{\bm G, \bm G'}(\bm k) & \equiv \braket{\Phi^{b,\chi}_{\bm k + \bm G} | \hat U | \Phi^{t,\chi'}_{\bm k + \bm G'}} \\
& = \frac{\sqrt{S_bS_t}}{S_{SC}} \sum_{\bm R_{\chi'}^t \in SC} e^{i(\bm G' - \bm  G) \cdot \bm R_{\chi'}^t} \\
& \times \sum_{\bm R_\chi^b} e^{i(\bm k + \bm G) \cdot (\bm R_{\chi'}^t - \bm R_\chi^b)} U(|\bm R_\chi^b - \bm R_{\chi'}^t|),
\end{aligned}
\end{equation}
with $S_{SC}$ the supercell area, and $S_b$ and $S_t$ the unit cell area of the bottom and top layer, respectively. Since $U(|\bm R_\chi^b - \bm R_{\chi'}^t|)$ decays exponentially, the sum over $\bm R_\chi^b$ can be truncated. We have checked convergence of the interlayer matrix elements for all results.

Putting everything together and writing the basis coefficients as $C^{i,\chi}_{\bm G} (\vec{k}) \equiv \braket{\Phi_{\bm k + \bm G}^{i,\chi}|\Psi_{\vec{k}}}$, the Schr\"odinger equation becomes
\begin{widetext}
\begin{alignat}{3}
\sum_{\chi'} \left[H_0^b(\bm k + \bm G)\right]_{\chi,\chi'} C^{b,\chi'}_{\bm G}(\bm k) & + \sum_{\bm G' \in BZ^{(t)},\chi'} U^{\chi,\chi'}_{\bm G, \bm G'}(\bm k) C^{t,\chi'}_{\bm G'} (\bm k) && = E_{\bm k} C^{b,\chi}_{\bm G} (\bm k), \label{schroddef2a} \\
\sum_{\chi'} \left[H_0^t(\bm k + \bm G)\right]_{\chi,\chi'} C^{t,\chi'}_{\bm G}(\bm k) & + \sum_{\bm G' \in BZ^{(b)},\chi'} U^{\chi',\chi}_{\bm G', \bm G}(\bm k)^*C^{b,\chi'}_{\bm G'} (\bm k) && = E_{\bm k} C^{t,\chi}_{\bm G} (\bm k), \label{schroddef2b}
\end{alignat}
\end{widetext}
for $\chi= A, B$ and $\bm G$ inside the BZ of the bottom (top) layer for the first (second) equation. For each $\bm k$, we obtain an eigenvalue equation whose solutions give the energy eigenvalues and eigenstates of the superstructure. The size of the matrix that has to be diagonalized is equal to $2(n_b + n_t)$, where $n_b$ $(n_t)$ is the number of reciprocal superlattice points that lie inside the BZ of the bottom (top) layer. Note that this corresponds with the number of atoms in the supercell of the respective layers.

As an example, consider AB-stacked bilayer graphene, for which $S_b=S_t=S_{SC}$ and only $\bm G = 0$ has to be considered for both layers. For the case where the $B$ sublattices lie on top of each other, we obtain the correct limit up to lowest order, including skew interlayer hopping \cite{trigonal},
\begin{equation}
\label{abham}
H(\vec{k}) =
\begin{pmatrix}
0 & \gamma_0 f(\bm k) & \gamma_4f^*(\bm k) & \gamma_4f(\bm k) \\
\gamma_0f^*(\bm k) & 0 & \gamma_4f(\bm k) & \gamma_1 \\
\gamma_4f(\bm k) & \gamma_4f^*(\bm k) & 0 & \gamma_0f^*(\bm k) \\
\gamma_4f^*(\bm k) & \gamma_1 & \gamma_0f(\bm k) & 0
\end{pmatrix},
\end{equation}
in the basis $\left( \ket{\Phi^{b,A}_{\bm k}}, \ket{\Phi^{b,B}_{\bm k}}, \ket{\Phi^{t,A}_{\bm k}}, \ket{\Phi^{t,B}_{\bm k}} \right)$. Note, however, that this model cannot distinguish the first-order interlayer skew hopping parameters $\gamma_3$ and $\gamma_4$, because we assumed that the interlayer coupling depends only on the distance between atoms and not on the environment of those atoms.

This model can be summarized as follows: In the absence of interlayer coupling, as a consequence of the increased periodicity of the superstructure, the bands of the individual layers fold onto the SBZ such that their original momenta are connected with a certain reciprocal superlattice vector that takes on the role of an additional band index in the SBZ. The interlayer coupling then induces coupling between states folded to the same SBZ momentum (as any other coupling is prohibited by translation symmetry). This generically leads to an avoided crossing. Since the energy reaches a local extremum in an anti-crossing, new peaks appear in the density of states.

Note that this approach can be used for all commensurate bilayer systems and the interlayer coupling can be taken into account up to arbitrary accuracy as long as the ansatz \eqref{g1var} is valid. Despite the fact that we only focus on bilayer structures, this model can easily be extended to multilayer systems. One downside, shared with tight-binding and first-principle calculations, is that calculations become computationally expensive for structures with large supercells. However, the above theory gives a better insight into the interlayer coupling mechanism, allowing for a straightforward low-energy approximation by reducing the amount of basis states. Moreover, perturbatively speaking, states at high energy can also be discarded. This allows one to introduce a suitable cut-off and systematically limit the number of basis states.

On the other hand, in a continuum model, the momentum space of the individual layers is given by the infinite plane, so that the basis is infinitely large. For this case, a cut-off is always necessary to limit the number of reciprocal superlattice vectors. Even though the interlayer matrix elements between states at $\bm k + \bm G$ and $\bm k + \bm G'$ become increasingly small as $|\bm G - \bm G'|$ becomes large compared to the inverse of the superlattice constant, as we will show later, the low-energy physics is extremely sensitive to perturbations, and failing to take into account intervalley scattering, for example, can possibly lead to inaccurate results for certain structures \cite{rotcont}.

Next, we scrutinize the case of twisted bilayer graphene to show that the model expounded in this section correctly describes all known features. We then proceed to novel faulted bilayer systems in which one of the layers is either triaxially or uniaxially strained, and we show that these systems exhibit a piezoelectric effect.

\section{Twisted bilayer graphene} \label{sec:Twisted bilayer graphene}

\begin{figure}
\centering
\includegraphics[width=8.5cm]{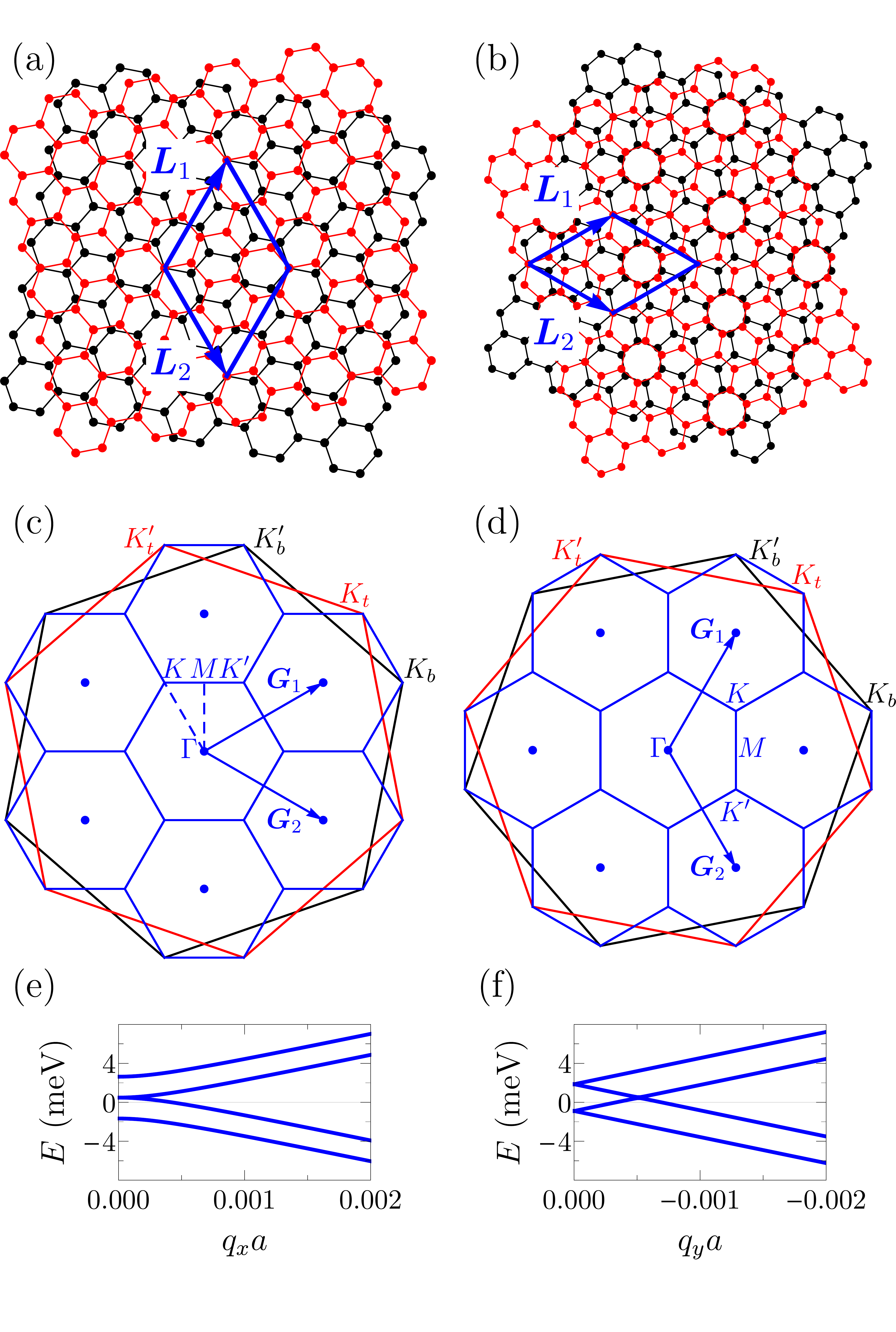}
\caption{(Color online) (a-b) Lattice of commensurate twisted bilayer with twist angle (a) $\theta=21.79^{\circ}$ $(m=2,n=1)$ and (b) $\theta=38.21^{\circ}$ $(m=4,n=1)$. The bottom layer is shown in black and the top layer in red. The primitive superlattice vectors $\bm L_{1,2}$ and the supercell are shown in blue. (c-d) Reciprocal lattice corresponding to (a) and (b), respectively. The large hexagons correspond to the BZ of the bottom and top layer. The smaller hexagons are the SBZs. The reciprocal superlattice vectors and the high-symmetry points are also shown. The path along which the bands in Fig.\ \ref{fig:DOSrot}(a) are plotted is indicated by the dashed line in (c). (e-f) Low-energy spectrum for (a) and (b), respectively, around the $K$ point with $\bm q = \bm k - \bm K$ along the $q_y=0$ (e) and $q_x=0$ (f) direction.}
\label{fig:tekenrot}
\end{figure}

One way to obtain a twisted bilayer is to first consider the case of AA-stacked bilayer graphene where the atoms of the individual layers are located above each other, and then rotate the bottom layer over an angle $-\theta/2$ and the top layer over an angle $\theta/2$ around an atomic site, as shown in Figs.\ \ref{fig:tekenrot}(a) and (b). Note that a rotation over $\theta=\pi/3$ results in AB-stacked bilayer graphene. The lattice vectors of the two layers are given by $\vec{a}_{1(2)}^b=aR(-\theta/2)\left(\sqrt{3}/2,\pm1/2\right)$ and $\vec{a}_{1(2)}^t=aR(\theta/2)\left(\sqrt{3}/2,\pm1/2\right)$, where $a = 2.46$ \AA\ is the graphene lattice constant. The intralayer Bloch Hamiltonian of both layers is then given by
\begin{equation}
\label{H0rot}
H_0^{b,t}(\vec{k}) = H_0 \left( R \left( \pm \theta/2 \right) \bm k \right),
\end{equation}
where $+$ $(-)$ corresponds to the bottom (top) layer, $H_0(\bm k)$ is the Bloch Hamiltonian of pristine graphene given in Eq.\ \eqref{H0}, and $R(\theta)$ is the rotation matrix for an anti-clockwise in-plane rotation over an angle $\theta$. In order to find a commensurate rotation, we demand that another pair of atoms in both layers overlap. If the pair at the rotation center belongs to sublattice $A$, the positions of the other pair can be written as $\bm R_{A^b} = m\bm a_1^b+n\bm a_2^b$ and $\bm R_{A^t} = m'\bm a_1^t+n'\bm a_2^t$, respectively, with $n,m,m',n'$ integers. Setting $m'=n$ and $n'=m$ leads to a generic nontrivial twist angle obtained from \cite{rotkosh}
\begin{equation}
\label{rothoek}
\cos \theta = \frac{1}{2}\frac{m^2+n^2+4mn}{m^2+n^2+mn},
\end{equation}
where $m$ and $n$ are coprime and $\theta$ is taken positive. A generic commensurate twist is therefore characterized by two coprime integers $m$ and $n$ with a twist angle given by Eq.\ \eqref{rothoek}. The primitive superlattice vectors can generically be chosen as
\begin{equation}
\label{Lrot}
\vec{L}_{1(2)} = \frac{L_c a}{2} \left(1, \pm \sqrt{3} \right),
\end{equation}
with $L_c=\sqrt{m^2+n^2+mn}$, implying that the Bravais lattice of a twisted bilayer is trigonal. The corresponding reciprocal superlattice vectors are given by
\begin{equation}
\label{Grot}
\vec{G}_{1(2)} = \frac{2\pi}{L_ca}\left(1,\pm\frac{1}{\sqrt{3}}\right).
\end{equation}
It follows that the SBZ, shown in Fig.\ \ref{fig:tekenrot}(c), has an area given by
\begin{align}
\label{brilopprot}
S_{BZ}^{SC} & = \frac{1}{2\sqrt{3}}\left(\frac{4\pi}{a}\right)^2 \frac{1}{L_c^2}, \\
S_{BZ}^b & = S_{BZ}^t = \frac{1}{2\sqrt{3}}\left(\frac{4\pi}{a}\right)^2,
\end{align}
so that the amount of reciprocal superlattice vectors that lie inside the BZ of the bottom and top layer is given by $n_b=n_t=L_c^2$.

\begin{figure}
\centering
\includegraphics[width=8.3cm]{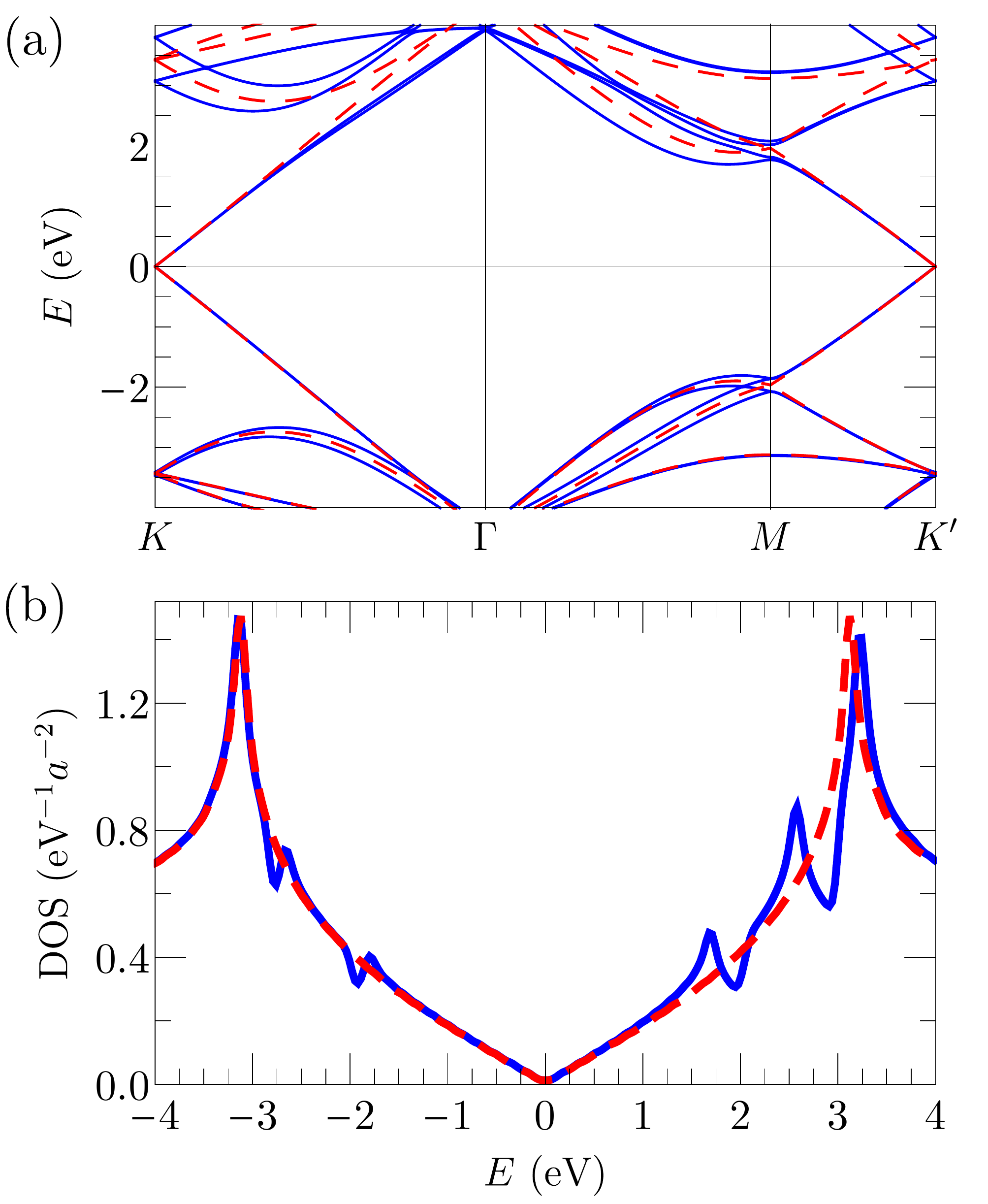}
\caption{(Color online) (a) Energy spectrum of twisted bilayer graphene with twist angle $\theta=21.79^{\circ}$ $(m=2,n=1)$ along the path in the SBZ indicated in Fig.\ \ref{fig:tekenrot}(c) with (solid, blue) and without (dashed, red) interlayer coupling. (b) Corresponding density of states of the band structure shown in (a).}
\label{fig:DOSrot}
\end{figure}

The energy bands of a commensurate twisted graphene bilayer with the smallest supercell $(m=2,n=1)$ is plotted along the high-symmetry directions of the SBZ in Fig.\ \ref{fig:DOSrot}(a). The results are in good agreement with the literature \cite{rotkosh}. In order to highlight the effect of the interlayer coupling, the band structure is also shown without interlayer coupling. As expected, the interlayer coupling generally lifts degeneracies leading to avoided crossings and it breaks the artificial symmetry between the valence and conduction band due to the inclusion of long-range interlayer hoppings. This is similar to the effect of the skew interlayer hopping parameter $\gamma_4$ in AB-stacked bilayer graphene \cite{ABAA1}. The corresponding density of states (DOS) is shown in Fig.\ \ref{fig:DOSrot}(b). Without interlayer coupling, the DOS is simply twice that of graphene, showing the two van Hove singularities at $E = \pm \gamma_0$. With interlayer coupling, the DOS becomes electron-hole asymmetric and obtains additional peaks due to anti-crossings in the band structure.
\begin{figure}
\centering
\includegraphics[width=8.3cm]{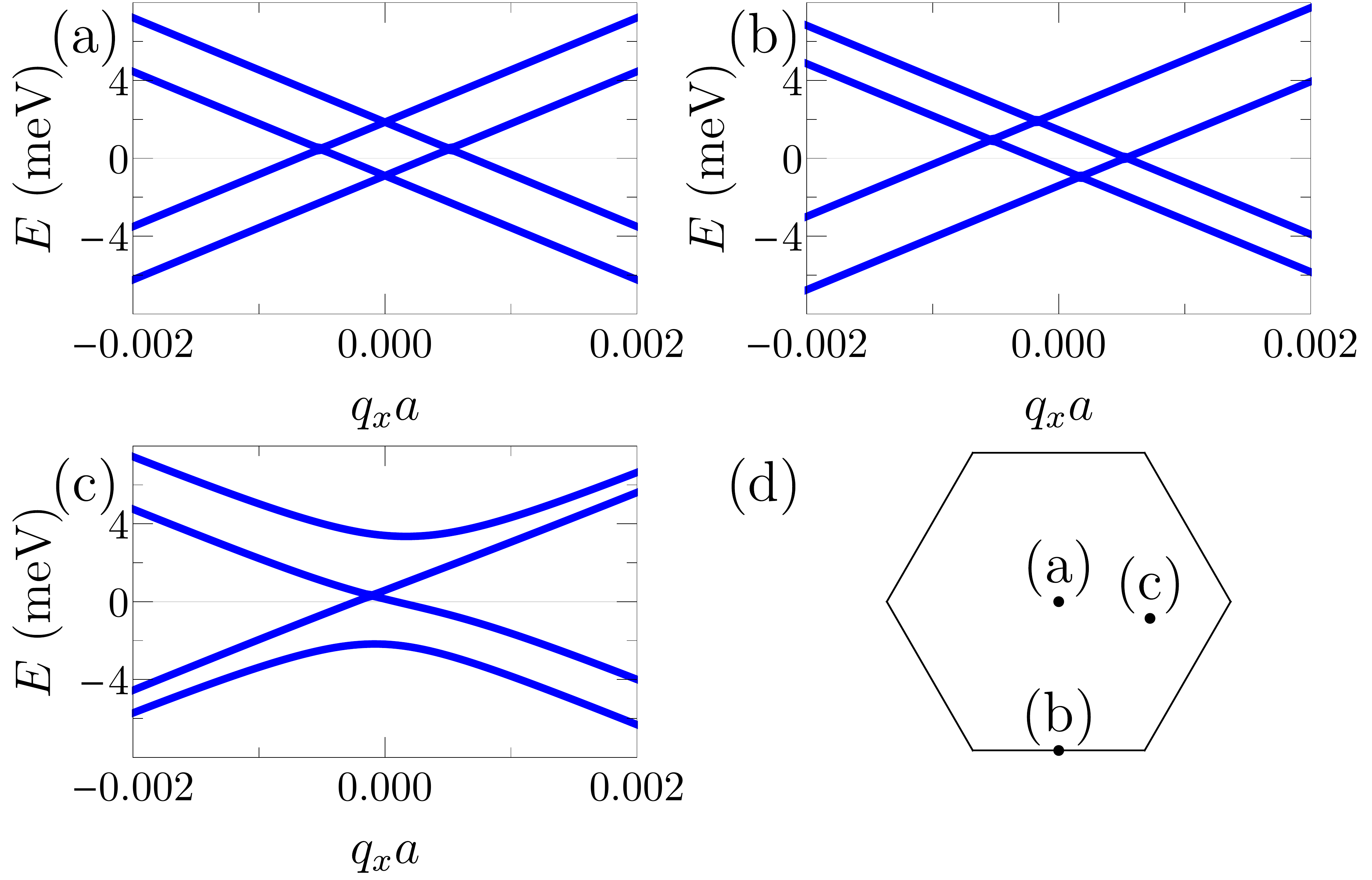}
\caption{(Color online) Low-energy spectrum of twisted bilayer with twist angle $\theta=21.79^{\circ}$ $(m=2,n=1)$ with $\bm q = \bm k - \bm K$ along the $q_y=0$ direction for a rotation around the center of a hexagon (a) and the middle of a bond (b), and along the appropriate $q_y\neq 0$ for a rotation around an arbitrary point (c). The center of rotation for each spectrum is shown in (d).}
\label{fig:lageEtwist}
\end{figure}
\begin{figure}
\centering
\includegraphics[width=7cm]{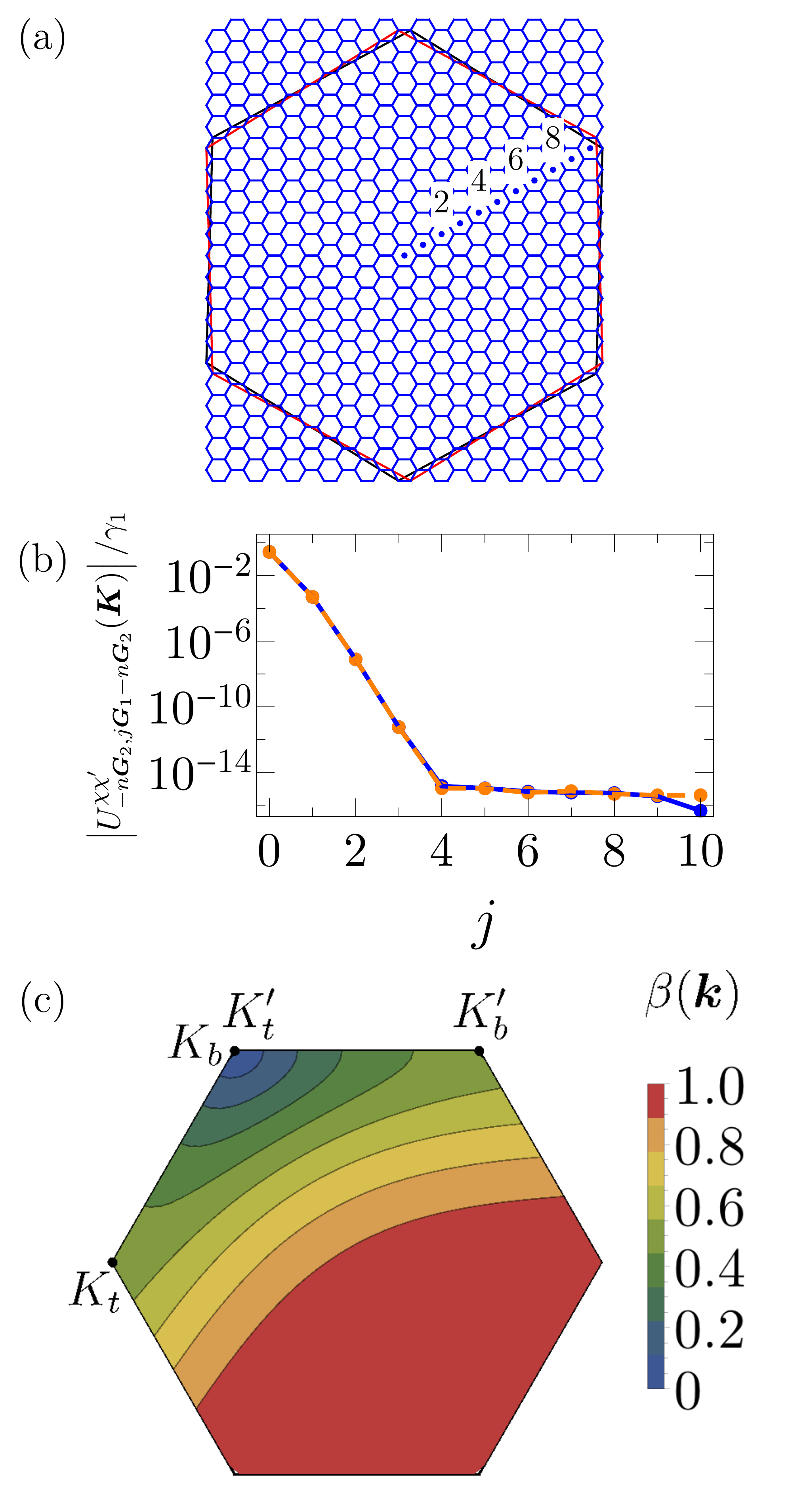}
\caption{(Color online) (a) Reciprocal lattice for a twisted bilayer with twist angle $\theta=3.15^{\circ}$ $(m=11,n=10)$. The large black and red hexagon indicate the BZ of the bottom and top layer, respectively. The smaller blue hexagons are the SBZs. (b) Interlayer coupling matrix element between states at $\bm K_b$ and $\bm K_b + j \bm G_1$ as a function of the integer $j$, between $A^b$ and $A^t$ (solid, blue) and $A^b$ and $B^t$ (dashed, orange). The reciprocal superlattice points $j \bm G_1$ are indicated in (a) by the points and labels. (c) Ratio of the interlayer coupling matrix elements $\beta (\bm k) \equiv |U^{A^bB^t}_{-n\vec{G}_2,n\vec{G}_1-n\vec{G}_2}(\vec{k})|/|U^{A^bA^t}_{-n\vec{G}_2,n\vec{G}_1-n\vec{G}_2}(\vec{k})|$ for the structure in Fig.\ \ref{fig:tekenrot}(a) throughout the SBZ.}
\label{fig:koppelingplot}
\end{figure}

In Fig.\ \ref{fig:tekenrot}(c), the reciprocal lattice is shown for the $(m=2,n=1)$ structure ($\theta = 21.79^\circ$). There are seven reciprocal superlattice vectors inside of the BZs of the bottom and top layer in agreement with $n_b=n_t=L_c^2 = 7$. We also see that $K_b$ and $K_t$, or $K_b'$ and $K_t'$, are folded to different momenta in the SBZ. There is, however, a reciprocal superlattice vector connecting $K_b$ and $K_t'$, as well as $K_t$ and $K_b'$, and therefore only intervalley coupling between the layers is present. Furthermore, from the lattice shown in Fig.\ \ref{fig:tekenrot}(a), we find that the only atoms that are located directly above each other come from the same sublattice as the rotation center. These types of structures have $C_3$ symmetry and are referred to in the literature as sublattice exchange (SE) odd \cite{mele,mele2}. It is therefore reasonable to expect that the low-energy spectrum is similar to that of AB-stacked bilayer graphene. This is indeed the case, as is shown in Fig.\ \ref{fig:tekenrot}(e), although the Dirac point is shifted up in energy by about $0.5$ meV due to long-range interlayer hopping.

Twisted bilayers determined by integers $m$ and $n$ for which $|m-n|$ is a multiple of three, lead to SE even structures. An example of such a system is shown in Fig.\ \ref{fig:tekenrot}(b). For these structures, the primitive and reciprocal superlattice vectors are given by
\begin{align}
\vec{L}_{1(2)} & = \frac{L_ca}{2\sqrt{3}}\left(\sqrt{3},\pm1\right), \label{Lrot2} \\
\vec{G}_{1(2)} & = \frac{2\pi\sqrt{3}}{L_ca}\left(\frac{1}{\sqrt{3}},\pm1\right), \label{Grot2}
\end{align}
giving $n_b = n_t = L_c^2/3$. In this case, there are no reciprocal superlattice vectors connecting $K_b$ and $K_t'$, or $K_t$ and $K_b'$, but $K_b$ and $K_t$ as well as $K_b'$ and $K_t'$ are folded on each other, so that there is only intravalley coupling between the layers for these structures. This is shown in Fig.\ \ref{fig:tekenrot}(d) taking $(m=4,n=1)$ ($\theta = 3.15^\circ$) as an example for which $n_b=n_t=7$. Furthermore, from Fig.\ \ref{fig:tekenrot}(b) we see that there are two atoms in the supercell, one of each sublattice, that coincide with atoms of the other layer. These types of structures have $C_6$ symmetry and are referred to as SE even \cite{mele,mele2}. Therefore, we expect that, in this case, the low-energy spectrum, shown in Fig.\ \ref{fig:tekenrot}(f), resembles that of AA-stacked bilayer graphene. The Dirac point is again shifted upwards in energy and the skew interlayer hopping results in a tiny gap at $|\bm q a|\approx-0.0005$ although it is not visible on the energy scale shown. These properties are generic for SE even structures for which $|m-n|$ is a multiple of three. The low-energy spectra of both classes are in good agreement with the literature \cite{mele,mele2}.

Different types of twisted structures are possible by rotating around other points of the lattice, such as the center of a hexagon, the middle of an in-plane bond, or an arbitrary point of the lattice, which are indicated in Fig.\ \ref{fig:lageEtwist}(d). A rotation around a hexagon center always results in an SE even structure with a low-energy spectrum resembling that of AA-stacked bilayer graphene, as shown in Fig.\ \ref{fig:lageEtwist}(a). The interlayer coupling mechanism differs depending on whether $|m-n|$ is a multiple of three or not, resulting in either intra- or intervalley coupling, respectively. Rotating around the middle of a bond results in a structure without $C_3$ symmetry belonging neither to the SE even or SE odd class. The corresponding low-energy spectrum is shown in Fig.\ \ref{fig:lageEtwist}(b). Note that the Dirac points are displaced from the $K$ point because of the absence of $C_3$ symmetry. This is also the case for the low-energy spectrum in Fig.\ \ref{fig:lageEtwist}(c) that corresponds to a rotation around an arbitrary point. Twisted structures with a different rotation center are all related to each other by a relative translation. Therefore, these results show some resemblance to the case of shifted bilayer graphene \cite{shifted}.

In Fig.\ \ref{fig:koppelingplot}(b), we show the absolute value of the interlayer coupling matrix elements between a Dirac point of the bottom layer and different states of the top layer for the SE odd system shown in Fig.\ \ref{fig:koppelingplot}(a). The matrix elements $U_{\bm G,\bm G'}$ decay very rapidly with increasing $|\bm G - \bm G'|$. This is clear from Eq.\ \eqref{matrixelemalgfin} since the sum becomes increasingly oscillatory, averaging out to zero. As such, a cut-off momentum can be introduced to limit the number of basis states, thus simplifying the problem. However, if we wish to construct a low-energy theory, we must be careful and consider all Dirac points even if they are separated by a large distance in momentum space since they are very susceptible to perturbations. Figure \ref{fig:koppelingplot}(b) also shows that the matrix elements between different sublattices are very similar.

Finally, we investigate the sublattice dependence of the matrix elements throughout the SBZ. In Fig.\ \ref{fig:koppelingplot}(c), we show the ratio of interlayer matrix elements between two different sublattices. We find that near the $K$ point onto which the $K_b$ and $K_t'$ points are folded, the interaction between the $A^b$ and $A^t$ sublattices dominates over the interaction between the $A^b$ and $B^t$ sublattices, whereas when moving away from this $K$ point, these interactions become equal in strength. The matrix elements between $A^b$ and $A^t$ are therefore the dominant interaction between the Dirac cones of different layers. This  is to be expected for an SE odd twisted bilayer, as these are the only sublattices located above each other.

\section{Triaxial strain} \label{sec:Triaxial strain}

\begin{figure}
\centering
\includegraphics[width=8.5cm]{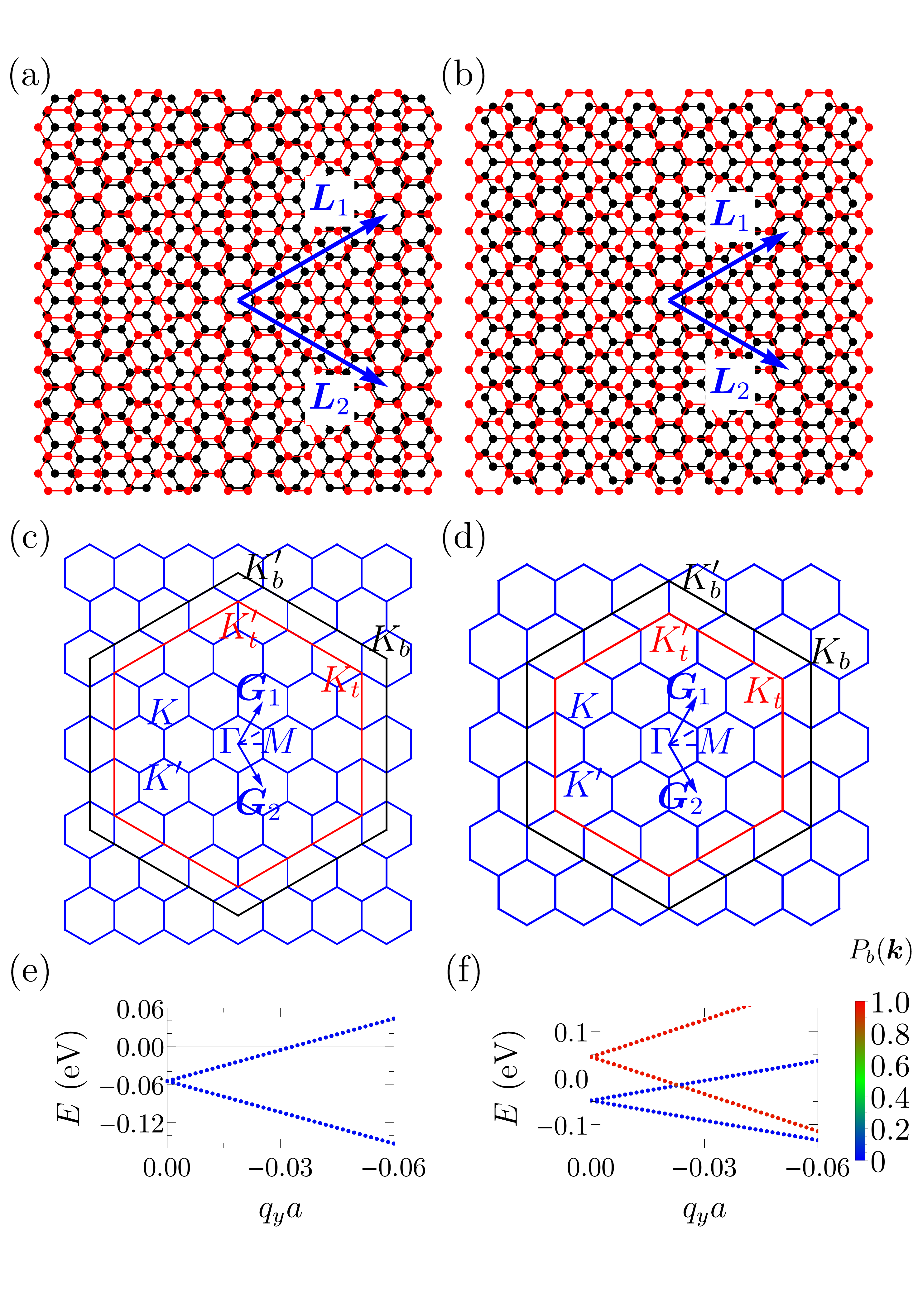}
\caption{(Color online) (a-b) Lattice of bilayer graphene for which the top layer is triaxially strained by (a) $\varepsilon_{tri}=20\%$ $(m=6,n=5)$ and (b) $\varepsilon_{tri}=25\%$ $(m=5,n=4)$. The bottom layer is shown in black and the top layer in red. The primitive superlattice vectors $\bm L_{1,2}$ and the supercell are shown in blue. (c-d) Reciprocal lattice corresponding to (a) and (b), respectively. The large hexagons correspond to the BZ of the bottom and top layer. The smaller hexagons are the SBZs. The reciprocal superlattice vectors and the high-symmetry points are also shown. The path along which the bands in Figs.\ \ref{fig:DOStri}(a-b) are plotted is indicated by the dashed line in (c-d), respectively. (e-f) Low-energy spectrum for (a) and (b), respectively, around the $K$ point with $\bm q = \bm k - \bm K$ along the $q_x=0$ direction. The colors of the bands show the fraction of the charge density localized on the bottom (unstrained) layer.}
\label{fig:tekentri}
\end{figure}

We start from AA-stacked bilayer graphene and apply tensile triaxial stress to the top layer while leaving the bottom layer unaltered, as shown in Figs.\ \ref{fig:tekentri}(a) and (b). The lattice vectors and Hamiltonian of the bottom layer are those of single-layer graphene. In order to describe the top layer we have to consider the strain tensor \cite{strain}
\begin{equation}
\label{straintensor}
\tilde \varepsilon (\varepsilon,\theta) = \varepsilon
\begin{pmatrix}
\cos^2\theta-\sigma\sin^2\theta & (1+\sigma)\sin\theta\cos\theta \\
(1+\sigma)\sin\theta\cos\theta & \sin^2\theta-\sigma\cos^2\theta
\end{pmatrix},
\end{equation}
where $\varepsilon$ is the stress-induced strain in the lattice, $\theta$ is the direction of the uniform stress with respect to the armchair direction and $\sigma=0.165$ is Poisson's ratio of graphite \cite{poisson}. Consequently any vector $\bm v$ is deformed up to first order to $\bm v' = \left(I_2 + \tilde \varepsilon \right) \bm v$, where $I_2$ is the identity matrix. Triaxial strain amounts to uniaxial strain along three axes that form angles of $2\pi/3$. The strain tensor for triaxial strain can thus be written as
\begin{equation}
\label{tristraintensor}
\begin{aligned}
I_2 + \tilde \varepsilon_{tri}(\varepsilon,\theta) &= \prod_{k=0}^2 \left[ I_2 + \tilde \varepsilon \left(\varepsilon,\theta + k\frac{2\pi}{3}\right) \right] \\
& \simeq I_2+\sum_{i=0}^2 \tilde \varepsilon \left(\varepsilon,\theta + k\frac{2\pi}{3}\right) \\
& = \left(1+\frac{3}{2}\varepsilon(1-\sigma)\right) I_2 \\
& \equiv \left(1+\varepsilon_{tri}\right)I_2,
\end{aligned}
\end{equation}
with $\varepsilon_{tri} = 3\varepsilon(1-\sigma)/2$. The lattice vectors of the strained top layer are given by
\begin{equation}
\label{topvectri}
\vec{a}_{1(2)}^t = \left(1+\varepsilon_{tri}\right)\vec{a}_{1(2)}^b = \frac{a}{2} \left(1+\varepsilon_{tri}\right)\left(\sqrt{3},\pm1\right).
\end{equation}
Since the strain is triaxial, the top layer is effectively a graphene layer with a larger lattice constant,  which can be up to 1.25 times larger since graphene can endure an in-plane strain up to 25\% [\onlinecite{sterk,sterk2}]. In turn, the strain changes the intralayer hopping parameter $\gamma_0$. Similar to Eq.\ \eqref{g1var}, the nearest-neighbor intralayer hopping of the strained layer is modeled with the ansatz \cite{strain}
\begin{equation}
\label{g0var}
\gamma_0(r) = \gamma_0 e^{-\alpha_s\left(r/a_0-1\right)},
\end{equation}
where $r$ is the distance between nearest neighbors and $\gamma_0 = 3.12$ eV is the intralayer hopping parameter of unstrained graphene. A commonly used value for the decay constant is $\alpha_s=3.37$, which agrees with predictions of the next nearest-neighbor hopping parameter and fits experimental results for $d\gamma_0(r)/dr$ [\onlinecite{g0var}]. The intralayer hopping parameter for the top layer becomes
\begin{equation}
\label{g0top}
\gamma_0^t(\varepsilon_{tri}) = \gamma_0e^{-\alpha_s\varepsilon_{tri}}.
\end{equation}
as a function of the triaxial strain $\varepsilon_{tri}$. The Hamiltonian of the top layer is then given by
\begin{equation}
\label{H0tri}
H_0^t(\vec{k}) =\frac{\gamma_0^t}{\gamma_0} H_0 \left( \left(1+\varepsilon_{tri}\right) \bm k \right).
\end{equation}
Commensurate structures require that the strain is a rational number:
\begin{equation}
\label{rationeeltri}
1+\varepsilon_{tri} = \frac{m}{n},
\end{equation}
with $m$ and $n$ coprime integers that characterize the structure. The primitive superlattice vectors and reciprocal superlattice vectors are given by
\begin{align}
\vec{L}_{1(2)} & = m\vec{a}_{1(2)}^b = n\vec{a}_{1(2)}^t = \frac{ma}{2} \left(\sqrt{3},\pm1\right), \label{Ltri} \\
\vec{G}_{1(2)} & = \frac{2\pi}{ma}\left(\frac{1}{\sqrt{3}},\pm1\right). \label{Gtri}
\end{align}
The SBZ is shown in Figs.\ \ref{fig:tekentri}(c) and (d) for two different structures as the small hexagon with area
\begin{equation}
\label{brilopptri}
S_{BZ} = \frac{1}{2\sqrt{3}}\left(\frac{4\pi}{ma}\right)^2.
\end{equation}
Note that the BZ area of the bottom and the top layer is given by
\begin{equation}
\label{bril12opptri}
S_{BZ}^b = \frac{1}{2\sqrt{3}}\left(\frac{4\pi}{a}\right)^2, \quad S_{BZ}^t = \frac{1}{2\sqrt{3}}\left(\frac{4\pi n}{ma}\right)^2,
\end{equation}
so that the number of reciprocal superlattice vectors inside the BZs of the bottom and top layer is given by $n_b=m^2$ and $n_t=n^2$, respectively.
\begin{figure}
\centering
\includegraphics[width=8.5cm]{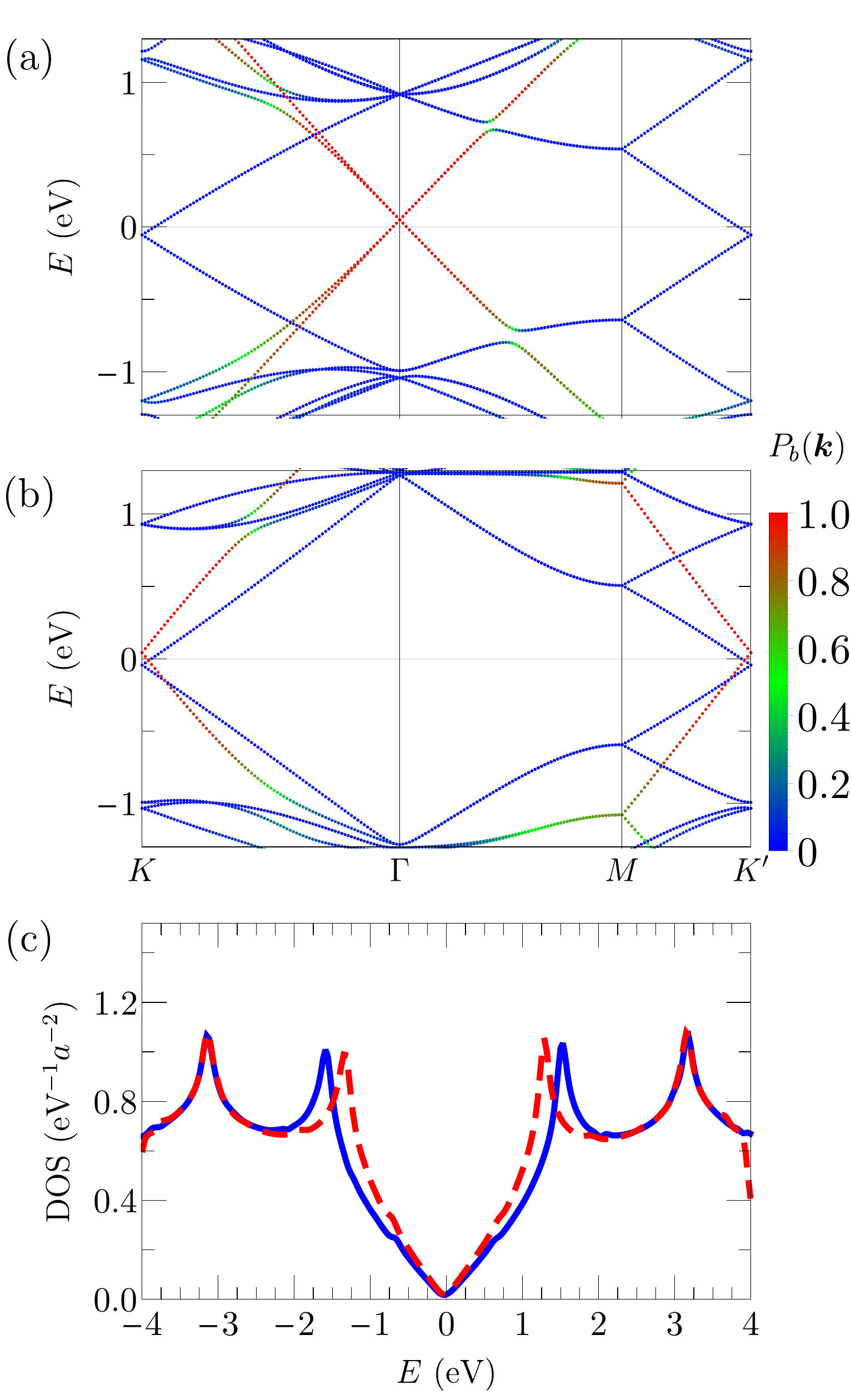}
\caption{(Color online) Energy spectrum of bilayer graphene for which the top layer is triaxially strained by (a) $\varepsilon_{tri}=20\%$ $(m=6,n=5)$ and (b) $\varepsilon_{tri}=25\%$ $(m=5,n=4)$ along the path in the SBZ indicated in Figs.\ \ref{fig:tekentri}(c-d). The colors of the bands show the fraction of the charge density localized on the bottom (unstrained) layer. (c) Density of states for (a) (solid, blue) and (b) (dashed, red).}
\label{fig:DOStri}
\end{figure}

For the $(m=6,n=5)$ structure, we see in Fig.\ \ref{fig:tekentri}(c) that $K_b$ and $K_b'$ are both folded to the $\Gamma$ point of the SBZ leading to a doubly degenerate cone. The Dirac points of the top layer, however, end up in the $K$ and $K'$ point of the SBZ, and the low-energy spectrum around these points consists of single cones. Since the Dirac points of the top and bottom layer are folded to different points, there is no low-energy interlayer coupling. This kind of structure occurs when either $m$ or $n$ is a multiple of three in which case the bottom or top cones are all folded to the SBZ center, respectively.

When neither $m$ nor $n$ is a multiple of three, the interlayer coupling is different. As an example, we consider the $(m=5,n=4)$ structure which is shown in Fig.\ \ref{fig:tekentri}(b). The corresponding reciprocal lattice is given in Fig.\ \ref{fig:tekentri}(d). In this case, there is intervalley coupling between the layers, where $K_t$ and $K_b'$ are folded to the $K$ point and $K_b$ and $K_t'$ are folded to the $K'$ point of the SBZ.

The difference between the two types of structures can also be understood by considering the lattices in Figs.\ \ref{fig:tekentri}(a) and (b). Structures for which neither $m$ nor $n$ is a multiple of three have an $A^b$ atom and a $B^t$ atom, as well as a $B^b$ atom and an $A^t$ atom, located above each other and we expect that the low-energy spectrum resembles that of AA-stacked bilayer graphene. Structures for which either $m$ or $n$ is a multiple of three do not have any atoms located directly above each other, and we expect a decoupling of the layers. Indeed, in this case, there is no interlayer interaction between the cones at all and the Dirac points in the resulting band structure are localized on one layer. This can be verified by calculating the layer polarization of the $i$th layer,
\begin{equation}
\label{polarisatie}
P_i(\vec{k}) = \sum_{\bm G \in BZ^{(i)}} \sum_{\chi=A^i,B^i} |C^{i,\chi}_{\bm G} (\bm k) |^2,
\end{equation}
for a given eigenstate, which is shown in Figs.\ \ref{fig:DOStri}(a) and (b) by the color of the bands.

The low-energy spectra of the structures with $\varepsilon_{tri}=20\%$ and $\varepsilon_{tri}=25\%$ are shown in Figs.\ \ref{fig:tekentri}(e) and (f), respectively. For the structure with $\varepsilon_{tri}=20\%$, the low-energy physics of the layers are decoupled since the Dirac cones from the unstrained layer are folded to the center of the SBZ, while the cones of the strained layer are folded to the $K$ and $K'$ point. Furthermore, the Dirac point of the strained top layer is shifted by about $50$ meV downwards in energy. In case $\varepsilon_{tri}=25\%$, the Dirac cones from different layers are folded on top of each other, but they remain largely localized within their respective layer, which can be seen in Fig.\ \ref{fig:tekentri}(f). This is mostly due to the large momentum separation between the original Dirac points which suppresses the interlayer coupling matrix element given in Eq.\ \eqref{matrixelemalgfin}. The cone associated with the strained layer is again shifted downwards in energy while the cone associated with the unstrained layer shifts upwards.

The full band structure of these structures is shown along high-symmetry directions in Figs.\ \ref{fig:DOStri}(a) and (b). There is a noticeable difference between them, since the Dirac points of the bottom unstrained layer are folded to the $\Gamma$ point for $\varepsilon_{tri}=20\%$, but not for $\varepsilon_{tri}=25\%$. Note that the cones from the strained layer have a smaller Fermi velocity than the cones coming from the unstrained layer due to the triaxial strain. This is clear since the Fermi velocity in the nearest-neighbor model of graphene is given by $v_F = \sqrt{3}\gamma_0a/(2\hbar)$ [\onlinecite{elecpropgraph}], and although the lattice constant of the strained layer increases linearly, the hopping parameter $\gamma_0^t$ decreases exponentially \cite{fermistrain}.

The electron-hole asymmetry of the band structure and the DOS, shown in Fig.\ \ref{fig:DOStri}(c), is again caused by long-range interlayer hopping. The two peaks in the DOS are due to the extrema of the bands in the $\Gamma$ point, where the peak at lower (higher) energy comes from the strained (unstrained) layer. Note that the peaks coming from the strained layer shift to lower energies with increasing strain because of the decrease in $\gamma_0^t$, while the peaks of the unstrained layer remain at $E = \pm \gamma_0$. Since the layers are largely decoupled, no prominent new peaks appear in the DOS. However, some avoided crossings appear around which the states are fully hybridized as shown in Figs.\ \ref{fig:DOStri}(a-b).

\begin{figure}[tb]
\centering
\includegraphics[width=7.5cm]{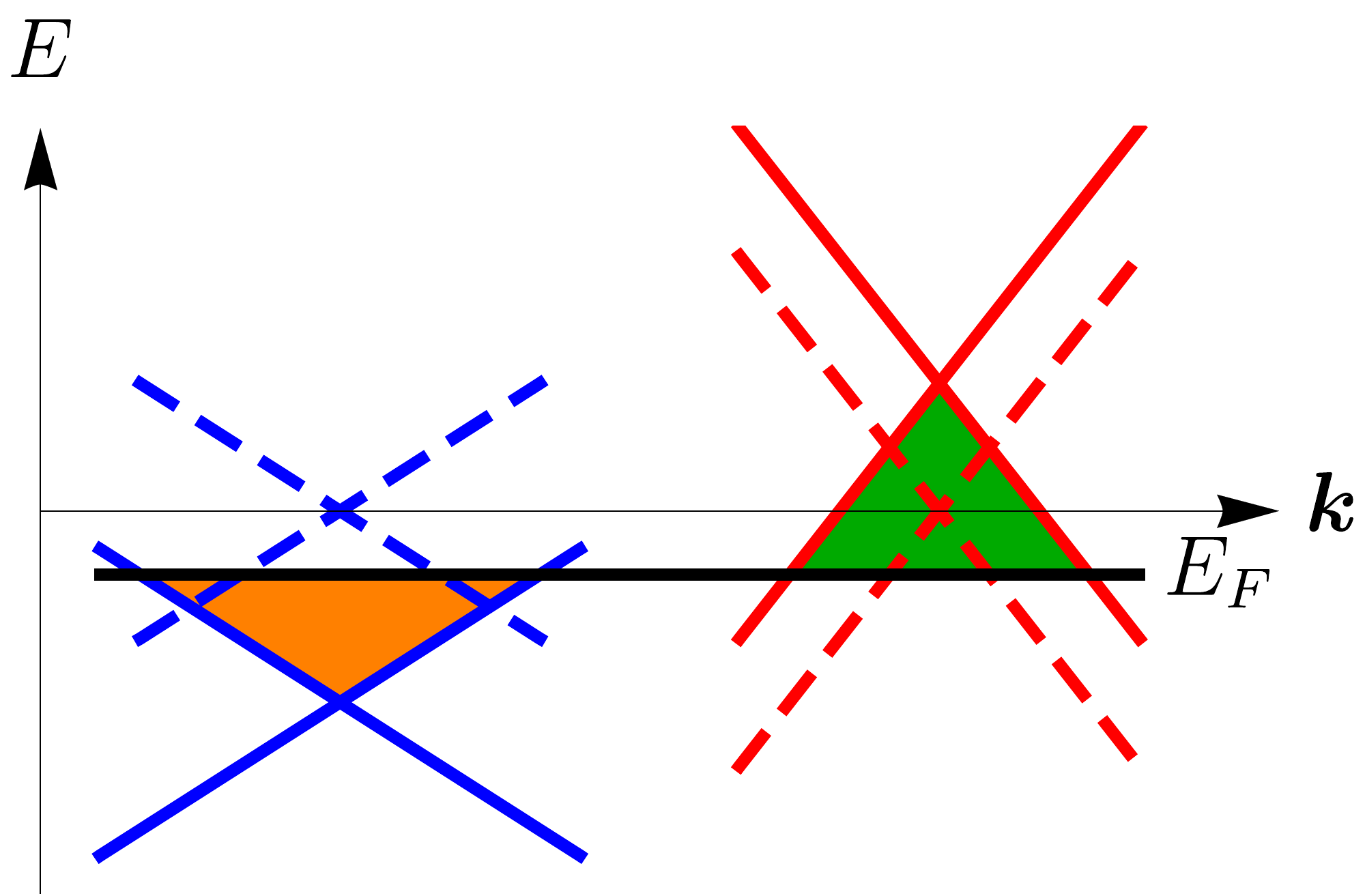}
\caption{(Color online) Schematic representation of the low-energy spectrum of bilayer graphene for which one layer is triaxially strained, with (solid) and without (dashed) interlayer coupling. The cones of the strained and unstrained layer are shown to the left (blue) and right (red), respectively. Electron (orange) and hole (green) concentrations are indicated by the filled triangles.}
\label{fig:schema}
\end{figure}

\begin{figure}
\centering
\includegraphics[width=8.3cm]{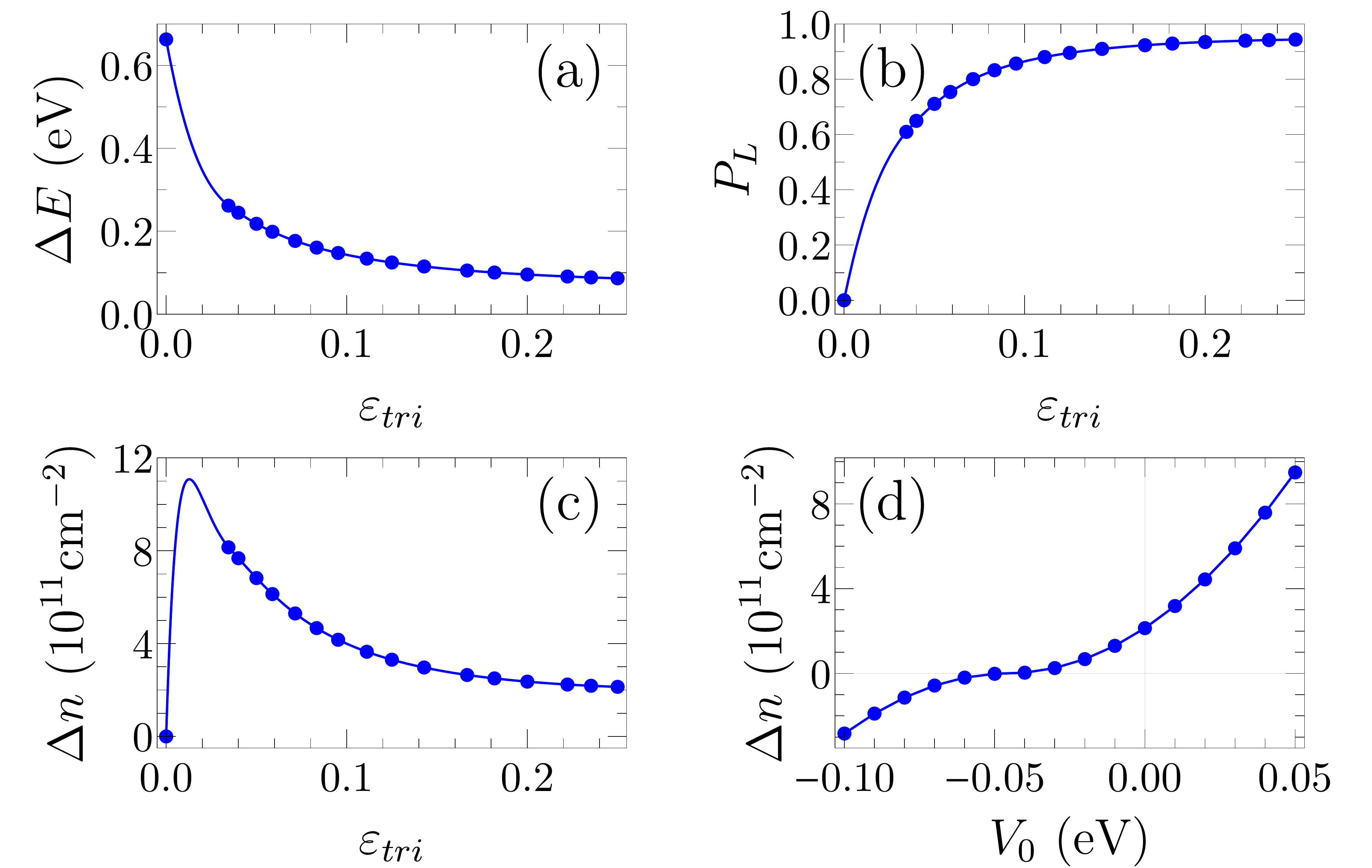}
\caption{(Color online) (a) Energy difference between the Dirac points of the two layers $\Delta E = E_D^b-E_D^t$ as a function of the triaxial strain $\varepsilon_{tri}$ on the top layer. (b) Layer polarization $P_L = P_b(\vec{D}_b)+P_t(\vec{D}_t)-1$ as a function of the strain. (c) Charge transfer from the bottom to the top layer as a function of the strain on the top layer, calculated from the interpolated curves in (a) and (b) and Eq.\ \eqref{transfer}. (d) Charge transfer for $\varepsilon_{tri}=25\%$ $(m=5,n=4)$ as a function of the interlayer bias potential $V_0$.}
\label{fig:concentratie}
\end{figure}

Due to the energy shift of the Dirac cones, schematically shown in Fig.\ \ref{fig:schema}, and the fact that they are localized mostly within one layer, charge is transferred between the cones and, therefore, between layers. Straining the top layer increases the carrier density of the top Dirac cone if the Fermi energy is kept constant, because the Fermi velocity is reduced. As such, electrons are transferred from the bottom to the top layer. The carrier concentration of a Dirac cone is given by
\begin{equation}
\label{conccone}
n(E_F) = \frac{1}{\pi}\left(\frac{E_D-E_F}{\hbar v_F}\right)^2,
\end{equation}
where $E_D$ is the energy of the Dirac point and $E_F$ the Fermi energy. Here we included the spin and valley degeneracy. The Fermi energy is then found by equating the carrier concentrations of the two cones:
\begin{equation}
\label{fermi}
E_F = \frac{v_F^tE_D^b+v_F^bE_D^t}{v_F^t+v_F^b},
\end{equation}
with $v_F^i$ and $E_D^i$ the Fermi velocity and the Dirac point energy of layer $i$. Plugging this into Eq.\ \eqref{conccone}, we obtain the charge transfer between the cones. We can relate this cone transfer to a layer transfer by taking into account the layer polarization from Eq.\ \eqref{polarisatie}. If $\bm D_i$ is the momentum of the Dirac point of the cones that originally came from the $i$th layer, only a fraction $P_b(\vec{D}_b)P_t(\vec{D}_t)$ of the cone transfer directly corresponds to charge transfer from the unstrained bottom layer to the strained top layer, while a fraction $(1-P_b(\vec{D}_b))(1-P_t(\vec{D}_t))$ is actually related to charge transfer from top to bottom. Their difference gives the total amount of charge transfer from the unstrained bottom layer to the strained top layer:
\begin{equation}
\label{transfer}
\Delta n = \frac{P_b(\vec{D}_b)+P_t(\vec{D}_t)-1}{\pi\hbar^2}\left(\frac{E_D^b-E_D^t}{v_F^b+v_F^t}\right)^2.
\end{equation}
In Fig.\ \ref{fig:concentratie}(c), $\Delta n$ is shown as a function of the strain. The charge transfer initially increases with the strain, then reaches a maximum and subsequently starts to decrease. We can understand this as follows: Initially, without strain, the energy difference $\Delta E=E_D^b-E_D^t$,  shown in Fig.\ \ref{fig:concentratie}(a), is maximal and equal to the value in AA bilayer graphene. However, the bands are fully hybridized, meaning that $P_L=P_b(\vec{D}_b)+P_t(\vec{D}_t)-1=0$ as is shown in Fig.\ \ref{fig:concentratie}(b), and in this case the charge transfer vanishes. With increasing strain, the interlayer coupling decreases and the Dirac cones become more localized in their respective layers which leads to an increase of the charge transfer. However, as the layers become increasingly decoupled, the energy difference between the Dirac points keeps decreasing and the charge transfer starts to decrease likewise after reaching a maximum. 

Since the charge transfer depends on the strain, this can be regarded as a piezoelectric effect. Furthermore, due to the separation of electrons and holes between the layers, this system can be a promising candidate for excitonic superfluidity \cite{super}. Note that these systems are non-centrosymmetric, as is required for the appearance of a piezoelectric effect\cite{centro}. The other aspect of the piezoelectric effect, the internal generation of strain resulting from an applied electric field, can not be studied with the present theory.

Finally, we considered the effect on the charge transfer of a uniform electric field that is applied perpendicular to the layers, which can be modeled by a layer bias potential, given by $V_0$ for the bottom layer and $-V_0$ for the top layer \cite{Eveld}. In Fig.\ \ref{fig:concentratie}(d), the charge transfer is shown as a function of $V_0$. We find that the electric field can be used to tune the charge transfer: increasing $V_0$ leads to an increase in the layer polarization and the energy difference of the Dirac points, enhancing the charge transfer. Decreasing $V_0$ initially decreases the energy difference, and hence the charge transfer, until it vanishes and the charge transfer changes sign. For the case of $\varepsilon_{tri}=25\%$, we find that the charge transfer changes sign for $V_0 = -0.05$ eV. This corresponds to an electric field strength of $E\approx3000$ kV/cm, which is much larger than what is experimentally achievable and therefore implies that the piezoelectric charge transfer is substantially larger than a charge transfer obtained by applying an electric field.
  
\begin{figure}
\centering
\includegraphics[width=8.3cm]{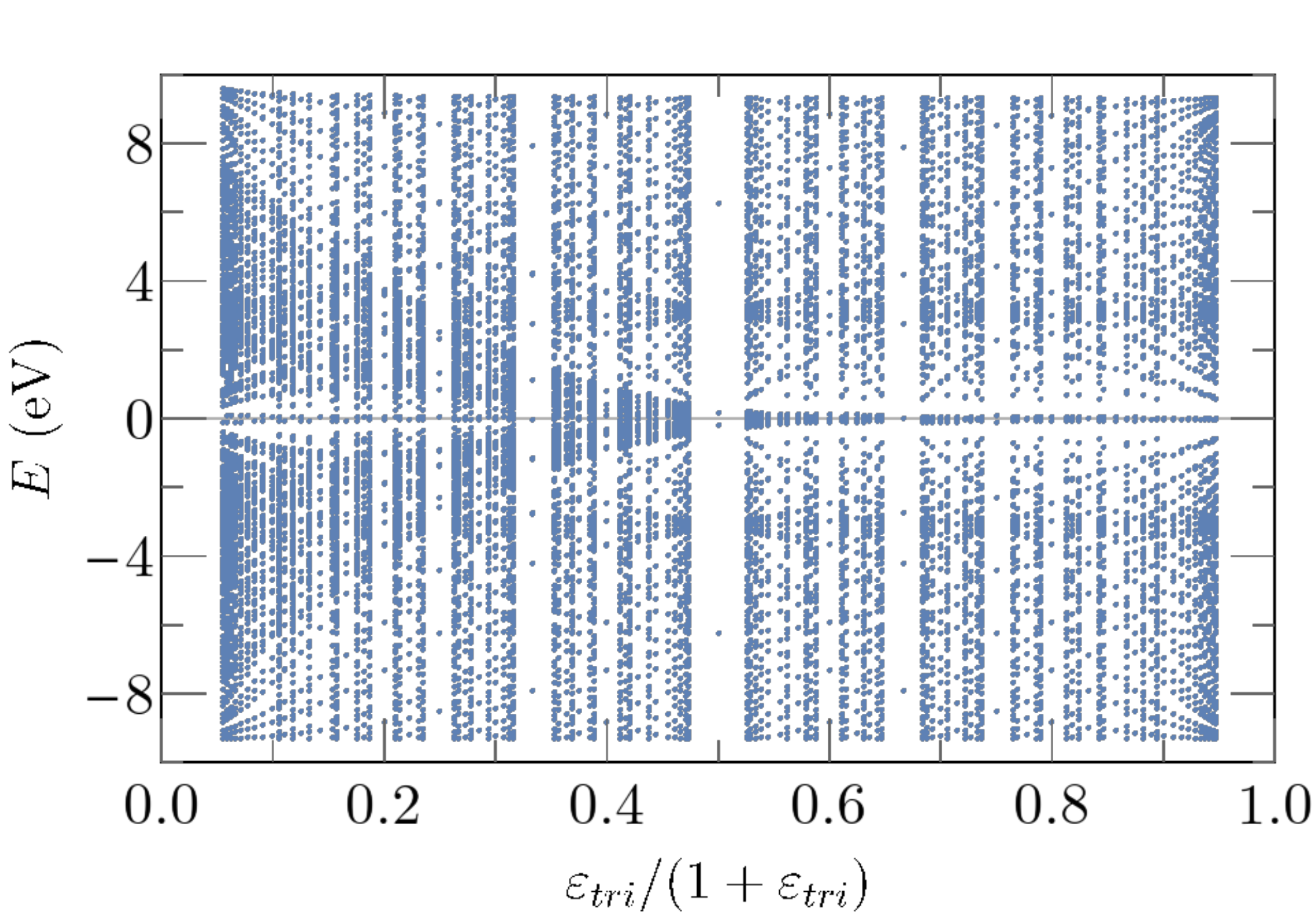}
\caption{(Color online) Self-similar spectrum of faulted bilayer graphene for which the top layer is triaxially strained, obtained by plotting the energy eigenvalues at the $K$ point of the SBZ as a function of $\varepsilon_{tri}/(1+\varepsilon_{tri})=1-n/m$. The plot is made for $m$ and $n$ coprime integers with $m$ ranging from 2 tot 19 and $n$ ranging from 1 to $m$.}
\label{fig:butterfly}
\end{figure}

In Fig.\ \ref{fig:butterfly}, the energy eigenvalues at the $K$ point of the SBZ are plotted as a function of $\varepsilon_{tri}/(1+\varepsilon_{tri})$. The resulting plot shows a self-similar structure reminiscent of the Hofstadter butterfly for Bloch electrons in a uniform perpendicular magnetic field \cite{hofstadter}. This kind of self-similarity is to be expected for commensurate superstructures, and was also found in other bilayer graphene systems with an applied magnetic field, both theoretically \cite{bilaagbutterfly,twistbutterfly} and experimentally \cite{expbutterfly}, and recently in corrugated carbon nanotubes \cite{nanosnake}.

\section{Uniaxial strain} \label{sec:Uniaxial strain}

\begin{figure}
\centering
\includegraphics[width=8.4cm]{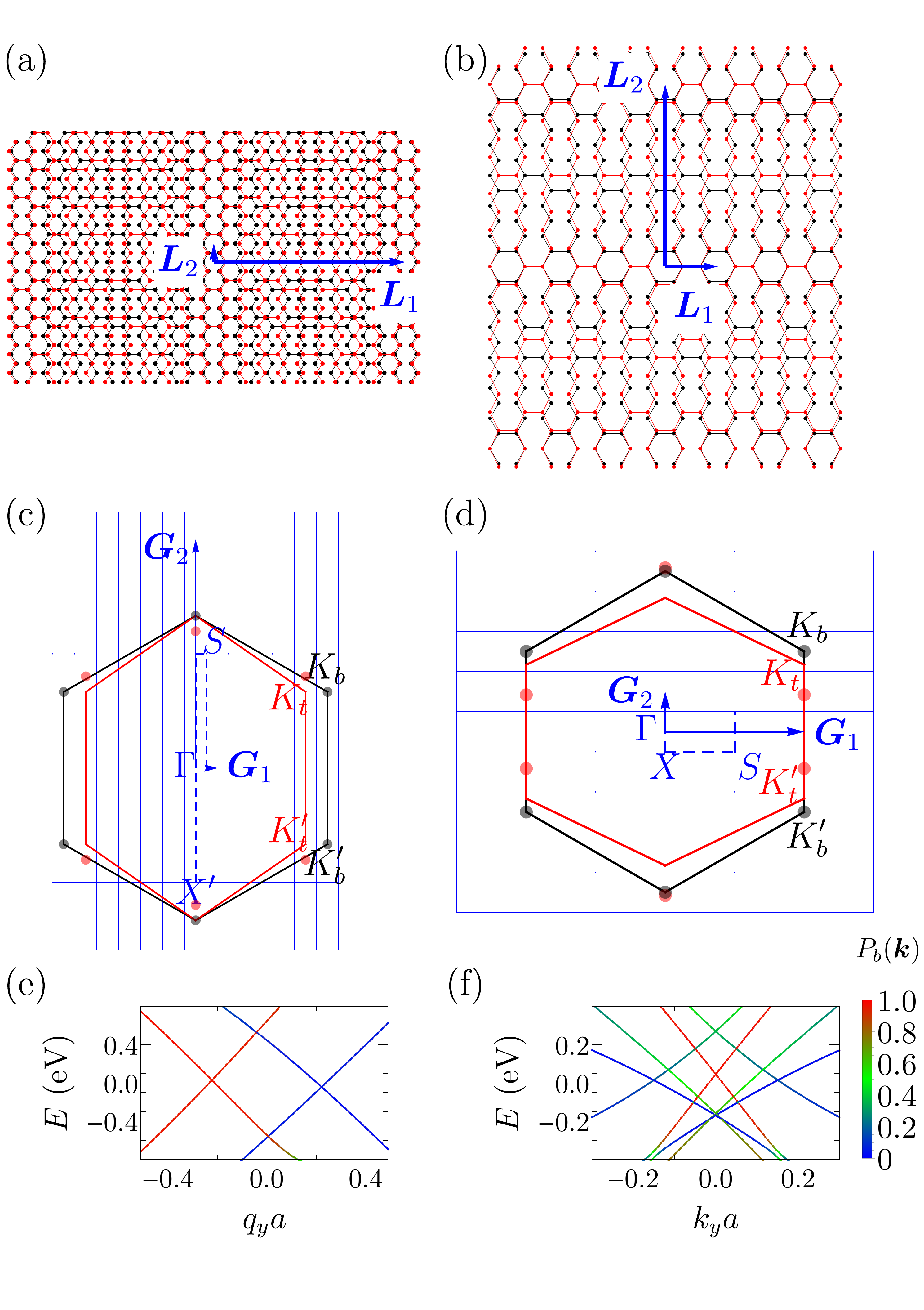}
\caption{(Color online) (a-b) Lattice of bilayer graphene for which the top layer is uniaxially strained with $\varepsilon=20\%$ $(m=6,n=5)$ in the (a) armchair and (b) zigzag direction. The bottom layer is shown in black and the top layer in red. The primitive superlattice vectors $\bm L_{1,2}$ and the supercell are shown in blue. (c-d) Reciprocal lattice corresponding to (a) and (b), respectively. The large hexagons correspond to the BZ of the bottom and top layer. The rectangular cells are the SBZs. The reciprocal superlattice vectors and high-symmetry points are also shown. The path along which the bands in Figs.\ \ref{fig:DOSuni}(a-b) are plotted is indicated by the dashed line in (c-d), respectively. The black and red dots are the Dirac points of the bottom and top layer, respectively. (e) Low-energy spectrum of (a) around the Dirac points with $\bm q = \bm k - \bm O$ along the $q_x=0$ direction. Here, $\bm O$ is the momentum halfway between the Dirac points of the two layers at positive $k_y$. (f) Low-energy spectrum of (b) along the $k_x=0$ direction. The colors of the bands show the fraction of the charge density localized on the bottom (unstrained) layer.}
\label{fig:tekenuni}
\end{figure}
\begin{figure}
\centering
\includegraphics[width=5.5cm]{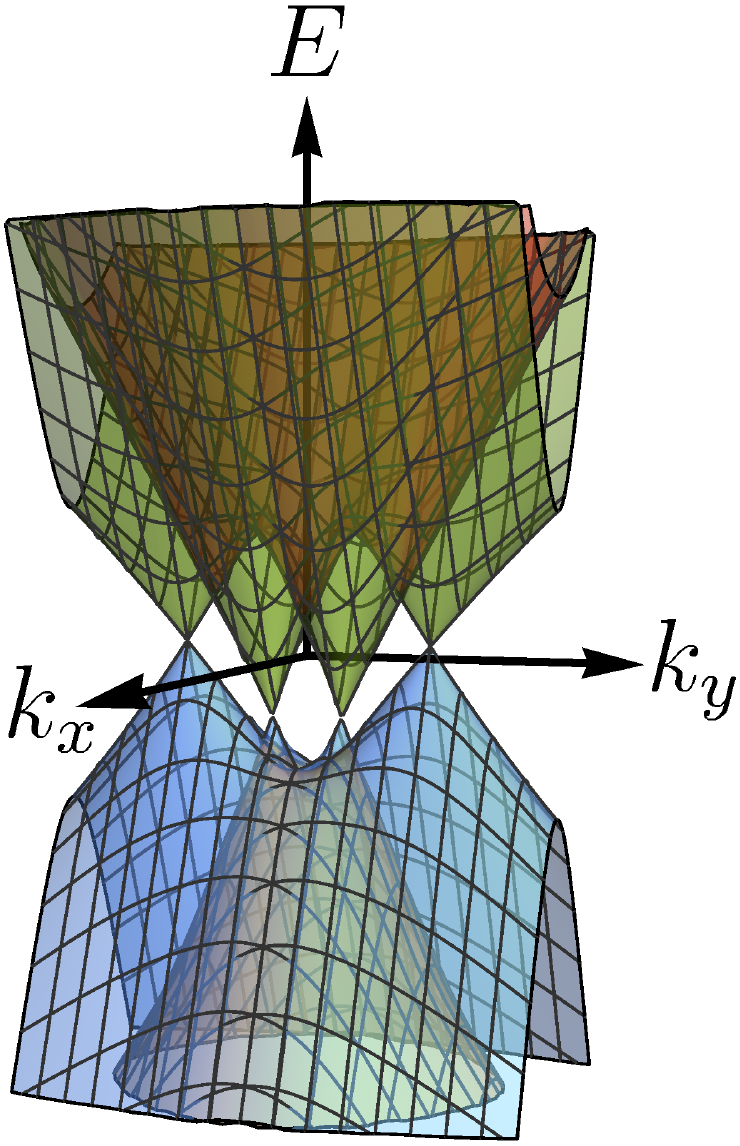}
\caption{(Color online) Four lowest energy bands of bilayer graphene for which the top layer is uniaxially strained with $\varepsilon=20\%$ $(m=6,n=5)$ in the zigzag direction. Note that all four Dirac points are located on the $k_y$-axis.}
\label{fig:3Dbanden}
\end{figure}
\begin{figure}
\centering
\includegraphics[width=8.5cm]{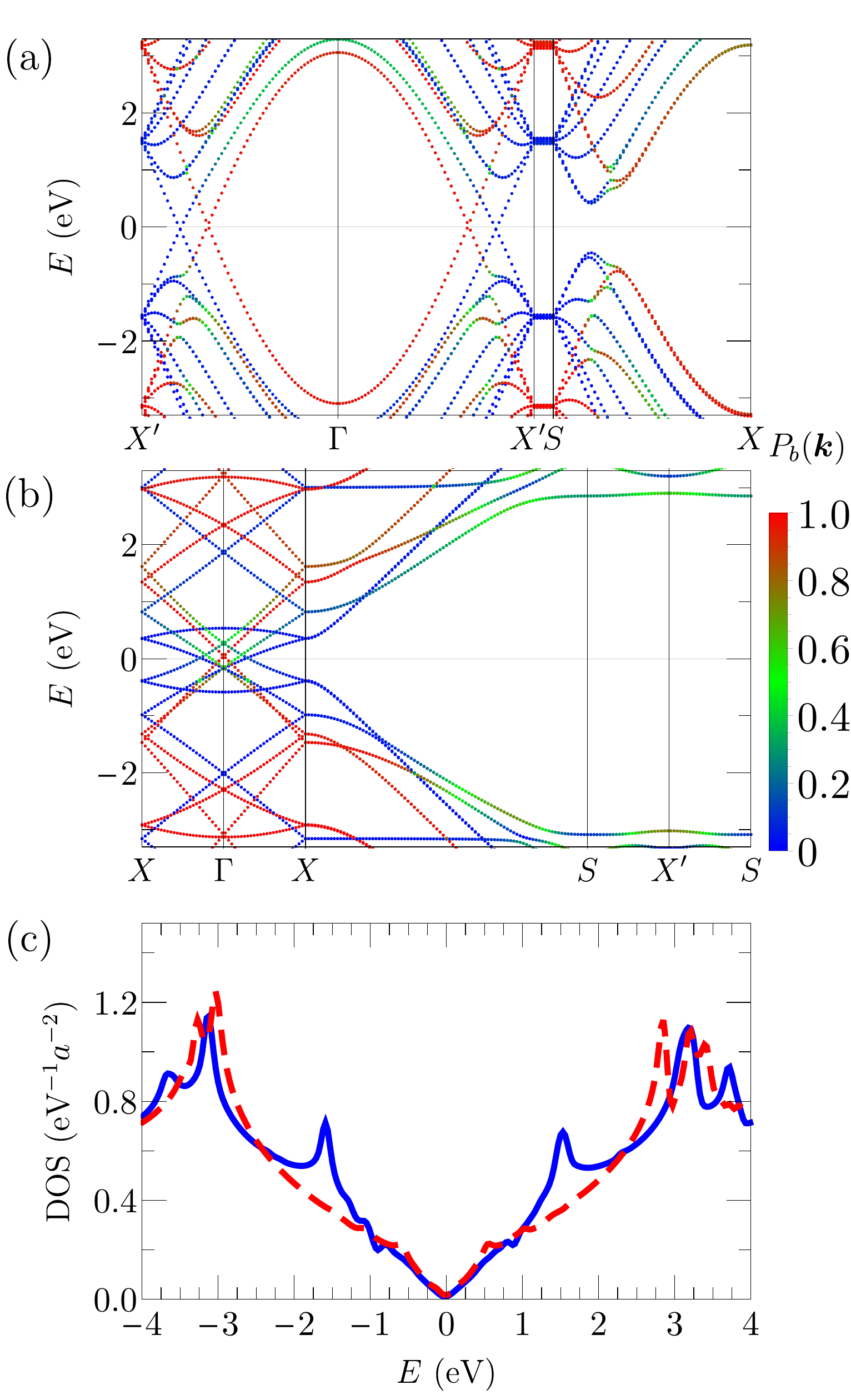}
\caption{(Color online) Energy spectrum of bilayer graphene for which the top layer is uniaxially strained with $\varepsilon=20\%$ $(m=6,n=5)$ in the (a) armchair and (b) zigzag direction, along the path in the SBZ shown in Figs.\ \ref{fig:tekenuni}(c-d). The colors of the bands show the fraction of the charge density localized on the bottom (unstrained) layer. (c) Density of states for (a) (solid, blue) and (b) (dashed, red).}
\label{fig:DOSuni}
\end{figure}

We consider the same system as discussed in the previous section where the top layer of the bilayer system is subjected to tensile stress but now we consider uniaxial stress instead. The lattice vectors and the Hamiltonian of the bottom layer, therefore, remain those of graphene. To study the strained top layer, we again use the strain tensor from Eq.\ \eqref{straintensor}. As an example, we consider two cases for the direction of the uniaxial strain: the armchair direction $(\theta=0$), shown in Fig.\ \ref{fig:tekenuni}(a), and the zigzag direction $(\theta=\pi/2)$, shown in Fig.\ \ref{fig:tekenuni}(b). The lattice vectors of the strained top layer are then given by
\begin{equation}
\label{topvecunia}
\vec{a}_{1(2)}^t = a\left((1+\varepsilon)\frac{\sqrt{3}}{2},\pm(1-\sigma\varepsilon)\frac{1}{2}\right),
\end{equation}
for the armchair direction, and
\begin{equation}
\label{topvecuniz}
\vec{a}_{1(2)}^t = a\left((1-\sigma\varepsilon)\frac{\sqrt{3}}{2},\pm(1+\varepsilon)\frac{1}{2}\right),
\end{equation}
for the zigzag direction. The lattice deformation again leads to a change in the intralayer hopping parameter $\gamma_0^t$ according to Eq.\ \eqref{g0var}. However, in this case $\gamma_0^t$ is not isotropic. The Hamiltonian of the top layer becomes
\begin{equation}
\label{H0uni}
H_0^t(\bm k) =
\begin{pmatrix}
0 & f_t(\bm k) \\ f_t^*(\bm k) & 0
\end{pmatrix},
\end{equation}
with $f_t(\bm k) = \sum_{j=1}^3 \gamma_{0j}^te^{i\bm k \cdot \bm \delta_j^t}$, where $\gamma_{0j}^t=\gamma_0 (\bm \delta_j^t)$. Here $\tilde{\varepsilon}$ is given by $\tilde{\varepsilon}(\varepsilon,0)$ or $\tilde{\varepsilon}(\varepsilon,\pi/2)$ for uniaxial strain in the armchair or zigzag direction, respectively, and $\vec{\delta}_j^t = (I_2+\tilde{\varepsilon}) \vec{\delta}_j^b$, are the three nearest-neighbor bond vectors of the top layer. Note that this implies that the Dirac points of the top layer move away from the high symmetry $K_t$ and $K_t'$ points in reciprocal space, since $C_3$ symmetry is broken. The location of the Dirac points of the top layer $\bm D_t$ follows from the condition $f_t(\vec{k})=0$, which gives
\begin{align}
D_{t,x} & = \left(2i+\frac{1}{2}+\eta\frac{1}{2}\right)\frac{2\pi}{\sqrt{3}a_{\xi}}, \\
D_{t,y} & = \left(\pm\frac{3}{\pi}\arccos\left(\eta\frac{\gamma_{01}}{2\gamma_{02}}\right)+6j\right)\frac{2\pi}{3a_{-\xi}},
\end{align}
with $\xi=\pm$ and where $\xi = +(-)$ if the stress is applied in the armchair (zigzag) direction. We also have $a_+=a(1+\varepsilon)$, $a_-=a(1-\sigma\varepsilon)$, $\eta=\pm$, and $i,j\in\mathbb{Z}$. The location of the Dirac points of the top layer is shown in Figs.\ \ref{fig:tekenuni}(c-d). Note that the labels of the high-symmetry points have changed because the SBZ is rectangular for uniaxial strain \cite{BZ}. In the remainder of  this section, we further assume $\sigma=0$ for simplicity.

The primitive superlattice vectors, respectively for the case of strain in the armchair and the zigzag direction, are given by
\begin{alignat}{3}
\vec{L}_1 & = a\left(m\sqrt{3},0\right) && \quad \text{and} \quad \vec{L}_2 && = a\left(0,m'\right), \\
\vec{L}_1 & = a\left(m'\sqrt{3},0\right) && \quad \text{and} \quad \vec{L}_2 && = a\left(0,m\right),
\end{alignat}
and the corresponding reciprocal superlattice vectors are, respectively, given by
\begin{alignat}{3}
\vec{G}_1 & = \frac{2\pi}{am}\left(\frac{1}{\sqrt{3}},0\right) && \quad \text{and} \quad \vec{G}_2 && = \frac{2\pi}{am'}\left(0,1\right), \\
\vec{G}_1 & = \frac{2\pi}{am'}\left(\frac{1}{\sqrt{3}},0\right) && \quad \text{and} \quad \vec{G}_2 && = \frac{2\pi}{am}\left(0,1\right),
\end{alignat}
with $m$ and $m'$ defined by
\begin{alignat}{3}
\label{rationeeluni}
1+\varepsilon & =  \frac{m}{n} && \quad \text{and} \quad 1-\sigma\varepsilon && = \frac{m'}{n'},
\end{alignat}
with $m$ and $n$ as well as $m'$ and $n'$ coprime integers. The reciprocal superlattice vectors define a rectangular SBZ for both cases, which is shown in Figs.\ \ref{fig:tekenuni}(c) and (d). The SBZ area becomes
\begin{equation}
\label{briloppuni}
S_{BZ} = \frac{4\pi^2}{\sqrt{3}ma^2},
\end{equation}
while the BZ area of the bottom and top layer is given by
\begin{equation}
\label{bril12oppuni}
S_{BZ}^b = \frac{1}{2\sqrt{3}}\left(\frac{4\pi}{a}\right)^2, \quad S_{BZ}^t = \frac{n}{2\sqrt{3}m}\left(\frac{4\pi}{a}\right)^2,
\end{equation}
so that $n_b=2m$ and $n_t=2n$.

As an example, we consider the $(m=6,n=5)$ structure for both strain in the armchair and zigzag direction. For strain in the armchair direction, both $K_b$ and $K_t$, and $K_b'$ and $K_t'$ are connected by reciprocal superlattice vectors, as shown in Fig.\ \ref{fig:tekenuni}(c). However, due to the fact that the Dirac points of the top layer move away from the $K_t$ and $K_t'$ points, there are no reciprocal superlattice vectors connecting the Dirac points of the two layers and the low-energy spectrum of both layers is decoupled. For strain in the zigzag direction, as can be seen from Fig.\ \ref{fig:tekenuni}(d), there are no reciprocal superlattice vectors connecting any corners of the BZs of the different layers and no reciprocal superlattice vectors connecting the Dirac points of the layers. However, $K_b$ and $K_b'$ are folded on each other, and, therefore, they connect the Dirac points of the bottom layer. This only occurs for structures for which $m$ is a multiple of three.

The low-energy spectra are shown in Figs.\ \ref{fig:tekenuni}(e) and (f). For uniaxial strain in the armchair direction there are two clearly separated Dirac points which are localized on different layers. The Dirac points of the bottom layer are shifted by about 50 meV upwards in energy and the Dirac points of the top layer by about the same amount downwards in energy. For uniaxial strain in the zigzag direction the Dirac points of the top layer are folded near the $\Gamma$ point, while those of the bottom layer are folded to the $\Gamma$ point. The Dirac points of the top layer are shifted upwards in energy by about 20 meV, while the Dirac points of the bottom layer shift downwards in energy by about 70 meV and are displaced from $\Gamma$ due to interlayer coupling. These cones consist partly of states associated with the bottom layer and partly of hybridized states. Therefore, we expect that this system also exhibits a piezoelectric effect. For clarity, the lowest four energy bands near $\Gamma$ are shown in Fig.\ \ref{fig:3Dbanden}.

The full energy spectra for the case of uniaxial strain in the armchair and zigzag direction along high-symmetry directions are shown in Figs.\ \ref{fig:DOSuni}(a) and (b), respectively. For both cases the interlayer coupling leads to the breaking of electron-hole symmetry and the hybridization of bands from different layers near avoided crossings. The density of states of the two structures are shown in Fig.\ \ref{fig:DOSuni}(c), which again shows the electron-hole asymmetry. Furthermore, the DOS for the case of uniaxial strain in the armchair direction shows extra peaks as compared to the case of uniaxial strain in the zigzag direction, stemming from highly degenerate bands between the $X'$ and $S$ points.

\section{Multilayer systems}
\label{sec:Multilayer}

The theory presented in Sec. \ref{sec:General theory} can be extended to commensurate multilayer systems, allowing the study of a wide variety of superstructures, e.g. combinations of twisted and strained layers or different layers with different twist angles or strain values. The total Hamiltonian for a general $N$-layer superstructure can be written as
\begin{equation}
\label{Nham}
\hat{H} = \sum_{i=1}^N\hat{H}^i_0+\sum_{j>i=1}^N\hat{U}^{ij},
\end{equation}
with $\hat{H}^i_0$ the intralayer Hamiltonian of layer $i$ and with $\hat{U}^{ij}$ the interlayer coupling between layer $i$ and $j$. Since the interlayer coupling $\hat{U}^{ij}$ has the combined periodicity of layers $i$ and $j$, this means, following the reasoning in Eq. \eqref{UGbew}, that this term only couples states $\ket{\Phi_{\vec{k}+\vec{G}}^{i,\chi}}$ of layer $i$ and $\ket{\Phi_{\vec{k}+\vec{G}'}^{j,\chi'}}$ of layer $j$ whose momenta differ by a reciprocal lattice vector $\vec{G}^{ij}$ of the combined system of layers $i$ and $j$, i.e. $\vec{G}-\vec{G}'=\vec{G}^{ij}$. The momentum $\vec{k}$ lies in the BZ of the total $N$-layer superstructure and $\vec{G}$ and $\vec{G}'$ are reciprocal lattice vectors of the total superstructure lying in the BZ of their corresponding layers. Similar to Eqs. \eqref{schroddef2a} and \eqref{schroddef2b}, we ultimately find
\begin{widetext}
\begin{equation}
\label{Nschrod}
\begin{split}
&\sum_{\chi'} \left[H_0^i(\bm k + \bm G)\right]_{\chi,\chi'} C^{i,\chi'}_{\bm G}(\bm k) + \sum_{\chi'}\Bigg(\sum_{j=i+1}^N\sum_{\substack{\bm G' \in BZ^{(j)}\\ \vec{G}-\vec{G}'=\vec{G}^{ij}}}U^{ij,\chi,\chi'}_{\bm G, \bm G'}(\bm k) + \sum_{j=1}^{i-1}\sum_{\substack{\bm G' \in BZ^{(j)}\\ \vec{G}-\vec{G}'=\vec{G}^{ij}}}U^{ji,\chi',\chi}_{\bm G', \bm G}(\bm k)^* \Bigg)C^{j,\chi'}_{\bm G'} (\bm k) = E_{\bm k} C^{i,\chi}_{\bm G} (\bm k),
\end{split}
\end{equation}
\end{widetext}
for each layer $i$, sublattice $\chi$ and reciprocal lattice vector of the total system $\vec{G}$ inside the BZ of layer $i$. These equations can again be solved to determine the energy spectrum and the eigenstates of the total superstructure.

\begin{figure}
\centering
\includegraphics[width=8.5cm]{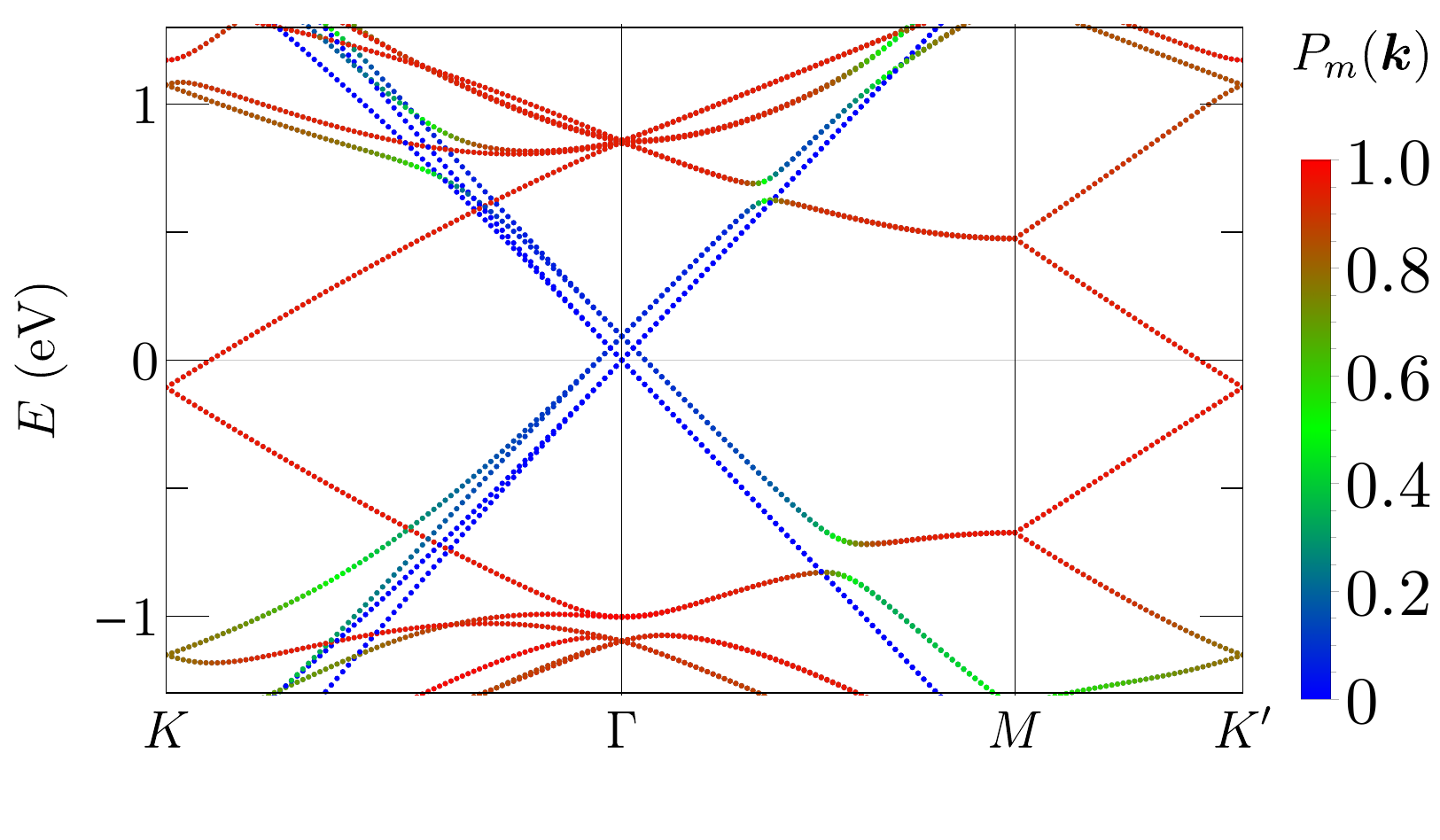}
\caption{(Color online) Energy spectrum of trilayer graphene for which the middle layer is triaxially strained by $\varepsilon_{tri}=20\%$. The colors of the bands show the fraction of the charge density localized on the middle (strained) layer $m$.}
\label{fig:trilaagtrimid}
\end{figure}

As an example, we consider a trilayer system in which, starting from AAA stacking, the middle layer is triaxially strained. The energy spectrum together with the layer polarization for this system is shown in Fig. \ref{fig:trilaagtrimid}. This spectrum is very similar to that of bilayer graphene for which the top layer is triaxially strained, shown in Fig. \ref{fig:DOStri}(a). The Dirac cones associated with the middle layer are located in the $K$ and $K'$ points, while the Dirac cones of the outer layers are weakly coupled with each other and are located in the $\Gamma$ point. Similar to the analysis performed for strained bilayer systems, we can infer that, because of the shifted and polarized Dirac cones, charge transfer now occurs from the outer layers to the middle layer.

\section{Summary and conclusion}
\label{sec:Conclusions}

In this paper, we constructed a theory for commensurate faulted bilayer graphene systems based on an expansion of the wave function in terms of the Bloch states of the separate layers. This theory accurately takes the interlayer coupling into account and can be used as a starting point to find the correct low-energy model. We have demonstrated the validity of the theory by first considering the well-studied case of twisted bilayer graphene and found that our results are in good agreement with the literature.

We then considered novel faulted bilayer systems in which triaxial or uniaxial stress is applied to one layer only. We found that, similar to the case of twisted bilayer graphene, there are two types of structures depending on the magnitude of the stress-induced strain. These two structures have a very different low-energy spectrum which can be understood from the interlayer coupling mechanism. For both types of strain, we found that the two layers can become decoupled with Dirac cones localized on separate layers. Because one of the layers is strained, there can be significant charge transfer from one layer to the other, in other words, these systems exhibit a piezoelectric effect. Moreover, we found that the strain-induced charge transfer can be tuned by applying a perpendicular electric field, which can enhance and reduce the charge transfer. Finally, we found a self-similar structure in the energy spectrum of bilayer graphene with triaxial stress applied to the top layer, similar to the Hofstadter butterfly.

Our approach can easily be modified to model other kinds of layered superstructures, for example a combination of twisting and straining one layer in a bilayer graphene system or even multilayer based superstructures, as we discussed in this paper. Furthermore, the existence of layer-separated electron and hole pockets in strained bilayer graphene could be promising for the study of excitonic superfluidity.

\section{Acknowledgments}

This work was supported by the Research Foundation-Flanders (FWO-Vl) through aspirant research grants to M.V.D.D., C.D.B., and B.V.D.


\begin{thebibliography}{10}

\bibitem{ontdekking}
K. S. Novoselov, A. K. Geim, S. V. Morozov, D. Jian, Y. Zhang, S. V. Dubonos, I. V. Grigorieva, and A. A. Firsov, Science \textbf{306}, 666 (2004).
\bibitem{elecpropgraph}
A. H. Castro Neto, N. M. R. Peres, K.S. Novoselov, and A. K. Geim, Rev. Mod. Phys. {\bf 81}, 109 (2009).
\bibitem{bilaag}
E. McCann and V. I. Fal'ko, Phys. Rev. Lett. \textbf{96}, 086805 (2006).
\bibitem{bernal}
J. D. Bernal, Proc. R. Soc. A \textbf{106}, 749 (1924).
\bibitem{ABAA1}
E. McCann and M. Koshino, Rep. Prog. Phys. \textbf{76}, 056503 (2013).
\bibitem{ABAA2}
A. Dyrdal and J. Barnas, Solid State Commun. \textbf{188}, 27 (2014).
\bibitem{lagen1}
C.-J. Shih,	A. Vijayaraghavan,	R. Krishnan,	R. Sharma,	J.-H. Han,	M.-H. Ham,	Z. Jin,	S. Lin,	G. L. C. Paulus,	N. F. Reuel, Q. Hua Wang, D. Blankschtein, and M. S. Strano, Nature Nanotech. \textbf{6}, 439 (2011).
\bibitem{lagen2}
H. Hiura, H. Miyazaki, and K. Tsukagoshi, Appl. Phys. Express \textbf{3}, 095101 (2010).
\bibitem{moire}
Z. Y. Rong and P. Kuiper, Phys. Rev. B \textbf{48}, 17427 (1993).
\bibitem{rotflat}
E. Su\'arez Morell, J. D. Correa, P. Vargas, M. Pancheco, and Z. Barticevic, Phys. Rev. B \textbf{82}, 121407(R) (2010).
\bibitem{rotrus}
A. O. Sboychakov, A. L. Rakhmanov, A. V. Rozhkov, and F. Nori, Phys. Rev. B \textbf{92}, 075402 (2015).
\bibitem{rotdft}
Z. Ni, L. Liu, Y. Wang, Z. Zheng, L.-J. Li, T. Yu, and Z. Shen, Phys. Rev. B \textbf{80}, 125404 (2009).
\bibitem{rotdft2}
G. T. de Laissardi\`ere, D. Mayou, and L. Magaud, Nano Lett \textbf{10}, 804 (2010).
\bibitem{rotoer}
J. M. B. Lopes dos Santos, N. M. R. Peres, and A. H. Castro Neto, Phys. Rev. Lett. \textbf{99}, 256802 (2007).
\bibitem{rottopol}
R. de Gail, M. O. Goerbig, F. Guinea, G. Montambaux, and A. H. Castro Neto, Phys. Rev. B \textbf{84}, 045436 (2011).
\bibitem{rotcont}
J. M. B. Lopes dos Santos, N. M. R. Peres, and A. H. Castro Neto, Phys. Rev. B \textbf{86}, 155449 (2012).
\bibitem{rotkosh}
P. Moon and M. Koshino, Phys. Rev. B \textbf{87}, 205404 (2013).
\bibitem{blochexp}
R. Bistritzer and A. H. MacDonald, Procl. Natl. Acad. Sci. USA \textbf{108}, 12233 (2011).
\bibitem{mele}
E. J. Mele, Phys. Rev. B \textbf{81}, 161405(R) (2010).
\bibitem{mele2}
E. J. Mele, J. Phys. D: Appl. Phys. \textbf{45}, 154004 (2012).
\bibitem{hbn1}
C. R. Dean, A. F. Young, I. Meric, C. Lee, L. Wang, S. Sorgenfrei, K. Watanabe, T. Taniguchi, P. Kim, K. L. Shepard, and J. Hone, Nat. Nanotechnol. \textbf{5}, 722 (2010).
\bibitem{hbn2}
J.  Xue,  J.  Sanchez-Yamagishi,  D.  Bulmash,  P.  Jacquod,  A. Deshpande, K. Watanabe, T. Taniguchi, P. Jarillo-Herrero, and B. J. LeRoy, Nat. Mater. \textbf{10}, 282 (2011).
\bibitem{hbn3}
B. Sachs, T. O. Wehling, M. I. Katsnelson, and A. I. Lichtenstein, Phys. Rev. B \textbf{84}, 195414 (2011).
\bibitem{hbn4}
M. Yankowitz, J. Xue, D. Cormode, J. D. Sanchez-Yamagishi, K. Watanabe, T. Taniguchi, P. Jarillo-Herrero, P. Jacquod, and B. J. LeRoy, Nat. Phys. \textbf{8}, 382 (2012).
\bibitem{uni1}
S.-M. Choi, S.-H. Jhi, and 	Y.-W. Son, Nano Lett. \textbf{10}, 3486 (2010).
\bibitem{uni2}
D. A. Gradinar, H. Schomerus, and V. I. Fal'ko, Phys. Rev. B \textbf{85}, 165429 (2012).
\bibitem{tbgdope1}
G. T. de Laissardi\`ere, O. M. Namarvar, D. Mayou, and L. Magaud, Phys. Rev. B \textbf{93}, 235135 (2016).
\bibitem{tbgdope2}
J. C. Rode, D. Smirnov, H. Schmidt, and R. J. Haug, 2D Mater. \textbf{3}, 035005 (2016).
\bibitem{graphdope}
M. T. Ong and E. J. Reed, ACS Nano \textbf{6}, 1387 (2012).
\bibitem{mos2}
 W. Wu, L. Wang, Y. Li, F. Zhang, L. Lin, S. Niu, D. Chenet, X. Zhang, Y. Hao, T. F. Heinz, J. Hone, and Z. L. Wang, Nature (London) \textbf{514}, 470 (2014).
\bibitem{mos22}
M. M. Alyoruk, Y. Aierken, D. Cakir, F. M. Peeters, and C. Sevik, J. Phys. Chem. C \textbf{119}, 23231 (2015).
\bibitem{graphnitr}
M. Zelisko, Y. Hanlumyuang, S. Yang, Y. Liu, C. Lei, J. Li, P. M. Ajayan, and P. Sharma, Nat. Commun. \textbf{5}, 4284 (2014).
\bibitem{tightbinding}
E. McCann and M. Koshino, Rep. Prog. Phys. \textbf{76}, 056503, (2013).
\bibitem{parameters}
B. Partoens and F. M. Peeters, Phys. Rev. B \textbf{74}, 075404 (2006).
\bibitem{elecpropgraf}
A. H. Castro Neto, F. Guinea, N. M. R. Peres, K. S. Novoselov, and A. K. Geim, Rev. Mod. Phys. \textbf{81}, 109 (2009).
\bibitem{trigonal}
E. McCann, D. S. L. Abergel, and V. I. Fal'ko, Solid State Commun. \textbf{143}, 110 (2007).
\bibitem{shifted}
M. Koshino, Phys. Rev. B \textbf{88}, 115409 (2013).
\bibitem{strain}
V. M. Perreira, A. H. Castro Neto, and N. M. R. Peres, Phys. Rev. B \textbf{80}, 045401 (2009).
\bibitem{poisson}
L. Blakslee, G. D. Proctor, E. J. Seldin, G. B. Stence, and T. Wen,  J. Appl. Phys. \textbf{41}, 3373 (1970).
\bibitem{sterk}
C. Lee, X. Wei, J. W. Kysar, and J. Hone, Science \textbf{321}, 385 (2008).
\bibitem{sterk2}
E. Cadelano, P. L. Palla, S. Giordano, and L. Colombo, Phys. Rev. Lett. \textbf{102}, 235502 (2009).
\bibitem{g0var}
R. O. Dillon, I. L. Spain, and J. W. McClure, J. Phys. Chem. Solids \textbf{38}, 635 (1977).
\bibitem{fermistrain}
F. de Juan, M. Sturla, and M. A. H. Vozmediano, Phys. Rev. Lett. \textbf{108}, 227205 (2012).
\bibitem{super}
M. Zarenia, A. Perali, D. Neilson, and F. M. Peeters, Sci. Rep. \textbf{4}, 7319 (2014).
\bibitem{centro}
J. F. Nye, \textit{Physical Properties of Crystals} (Oxford University Press, Oxford, 1985).
\bibitem{Eveld}
E. V. Castro, K. S. Novoselov, S. V. Morozov, N. M. R. Peres, J. M. B. Lopes dos Santos, J. Nilsson, F. Guinea, A. K. Geim, and A. H. Castro Neto, Phys. Rev. Lett. \textbf{99}, 216802 (2007).
\bibitem{hofstadter}
D. R. Hofstadter, Phys. Rev. B \textbf{14}, 2239 (1976).
\bibitem{bilaagbutterfly}
N. Nemec and G. Cuniberti, Phys. Rev. B \textbf{75}, 201404(R) (2007).
\bibitem{twistbutterfly}
R. Bistritzer and A. H. MacDonald, Phys. Rev. B \textbf{84}, 035440 (2011).
\bibitem{expbutterfly}
C. R. Dean, L. Wang, P. Maher, C. Forsythe,	F. Ghahari, Y. Gao, J. Katoch, M. Ishigami, P. Moon, M. Koshino, T. Taniguchi, K. Watanabe, K. L. Shepard, J. Hone, and P. Kim, Nature (London) \textbf{497}, 598 (2013).
\bibitem{nanosnake}
P. Gentile, M. Cuoco, and C. Ortix, Phys. Rev. Lett. \textbf{115}, 256801 (2015).
\bibitem{BZ}
W. Setyawan and S. Curtarolo, Comput. Mater. Sci. \textbf{49}, 299 (2010).

\end{thebibliography}
\end{document}